\documentclass{aastex63}
\usepackage{booktabs}
\usepackage{multirow}
\usepackage{bm}

\received{XXX}
\revised{XXX}
\accepted{XXX}
\submitjournal{AJ}

\shortauthors{Wang et al.}
\graphicspath{{./}{figures/}}

\begin{document}

\title{Follow-up photometry in another band benefits reducing \emph{Kepler}'s false positive rates}

\author[0000-0003-3015-6455]{Mu-Tian Wang}
\affiliation{School of Astronomy and Space Science,
                Nanjing University,
                Nanjing 210023, China.}
\affiliation{Key Laboratory of Modern Astronomy and Astrophysics, Ministry of Education, Nanjing, 210023, People’s Republic of China}

\author[0000-0001-5162-1753]{Hui-Gen Liu}
\affiliation{School of Astronomy and Space Science,
                Nanjing University,
                Nanjing 210023, China.}
\affiliation{Key Laboratory of Modern Astronomy and Astrophysics, Ministry of Education, Nanjing, 210023, People’s Republic of China}

\author[0000-0002-8775-4387]{Jiapeng Zhu}
\affiliation{School of Astronomy and Space Science,
                Nanjing University,
                Nanjing 210023, China.}
\affiliation{Key Laboratory of Modern Astronomy and Astrophysics, Ministry of Education, Nanjing, 210023, People’s Republic of China}

\author[0000-0003-1680-2940]{Ji-Lin Zhou}
\affiliation{School of Astronomy and Space Science,
                Nanjing University,
                Nanjing 210023, China.}
\affiliation{Key Laboratory of Modern Astronomy and Astrophysics, Ministry of Education, Nanjing, 210023, People’s Republic of China}

\begin{abstract}

\emph{Kepler} Mission's single-band photometry suffers from astrophysical false positives, the most common of background eclipsing binaries (BEBs) and companion transiting planets (CTPs). Multi-color photometry can reveal the color-dependent depth feature of false positives and thus exclude them. In this work, we aim to estimate the fraction of false positives that are unable to be classified by \emph{Kepler} alone but can be identified with their color-dependent depth feature if a reference band ($z$, $K_s$ and TESS) were adopted in follow-up observation. We build up physics-based blend models to simulate multi-band signals of false positives. Nearly 65-95\% of the BEBs and more than 80\% of the CTPs that host a Jupiter-size planet will show detectable depth variations if the reference band can achieve a \emph{Kepler}-like precision. $K_s$ band is most effective in eliminating BEBs exhibiting any depth sizes, while $z$ and TESS band prefer to identify giant candidates and their identification rates are more sensitive to photometric precision. Provided the radius distribution of planets transiting the secondary star in binary systems, we derive formalism to calculate the overall identification rate for CTPs. By comparing the likelihood distribution of the double-band depth ratio for BEB and planet models, we calculate the false positive probability (FPP) for typical \emph{Kepler} candidates. Additionally, we show that the FPP calculation helps distinguish the planet candidate's host star in an unresolved binary system. The analysis framework of this paper can be easily adapted to predict the multi-color photometry yield for other transit surveys, especially for TESS.

\end{abstract}


\section{Introduction}
 \label{sec:intro}

 \emph{Kepler} \citep{2010Sci...327..977B} is a space mission dedicated to search for periodic dimming in a star's brightness as evidence of planetary transits. However, non-planetary astrophysical objects can also produce transit-like light curves \citep{Brown2003}. These objects are called false positives (FPs). One common source of \emph{Kepler} FPs is a pair of background eclipsing binary (BEB) co-aligning with the \emph{Kepler} target star, in which the eclipse is diluted by the foreground star's flux and thus appear as the transit of a planet-class object. BEBs are a major source of FPs at early \emph{Kepler} data release, nearly accounting for 40\% of the Kepler Objects of Interests (KOIs) discovered in lower galactic latitude region \citep{Batalha2010,Batalha2013,Borucki2011,Bryson2010}. Besides, BEBs can easily mimic the transits of earth-sized candidates due to the large brightness difference between foreground and background components \citep{Fressin2013}. Another frequent source of FPs is planets transiting in binary systems. The planet radius inferred from the contaminated light curves will be miscalculated in absent knowledge about the stellar companion. Nearly 30\% of planet candidates' host stars have nearby stellar companions within 4 arcseconds \citep{Law2014,Ziegler2016,Furlan2017,Ziegler2018}, which is the one pixel-scale of \emph{Kepler}. If the planet is transiting the brighter target star, the flux contamination by the companion is at most 50\%. However, suppose the planet is transiting the much fainter companion star, termed companion transiting planets (CTPs). In that case, the dilution could be much higher than 50\%, and the planet's radius interpreted from diluted transit depths could be severely underestimated, leading to wrong planet classifications and affect occurrence rate study \citep{Borucki2011,Ziegler2016,Wang2018,Teske2018}. For example, the occurrence rate of Earth-size planets could be overestimated by tens of percents if not accounting for the binarity of KOI host star \citep{Bouma2018}. Given the ubiquity of these two kinds of FPs, separating FPs from genuine (and correctly classified) planets is crucial for transit survey like \emph{Kepler}, whose main objects are Earth-size planets in habitable zones.

 One way to separate these FPs from genuine planets is to measure the mass of transiting body through radial velocity (RV) measurements. However, RV observations are primarily based on several premium ground-based facilities that are of high demand and are best targeted for planet candidates with high mass from bright host stars, which are not typical for the ensemble of \emph{Kepler} candidates \citep{Borucki2011,Batalha2010}. Moreover, directly conducting RV observations to identify BEBs and CTPs could be problematic in that (1) the spectrum of background eclipsing binary can be concealed by the contamination of foreground star and (2) the RV residuals of stellar companions cannot be easily corrected without knowing the binarity of system. Thus, the conventional RV measurements are not the optimal way to comprehensively vet \emph{Kepler} candidates.
 
 Instead of directly measuring the planet's mass, \emph{statistical validation} is proposed to validate the transit candidates. \emph{Statistical validation} aims to argue that the transit signal is more likely to be created by planet transits than any other foreseeable FP scenarios. The prototype of this procedure, \texttt{BLENDER}, was first embodied and employed by \emph{Kepler} team \citep{Torres2005,Torres2011}. More recently, \cite{Morton2012}, and \cite{Diaz2014} develop more efficient tools to evaluate the candidates' false positive probability (FPP) under Bayesian framework. Most notably, the \texttt{VESPA} has successfully validated over a thousand of \emph{Kepler} candidates \citep{Morton2016}. A commonality of these methods is that any available ancillary information of the candidate (e.g. contrast curve, RVs, multi-color transit photometry, etc.) can be utilized to further constrain and clear the stellar parameter space of FPs that fail to reproduce the observed light curves. When the FPP is below the defined threshold, the candidate can be safely validated.

 Using multi-color photometry to vet transit candidates is based on the fact that blends containing stars with different colors will project color-dependent depth signature on different observing bandpasses \citep{Rosenblatt1971,Tingley2004}. As the color-difference between different components increases, the difference between mutil-band transit depths will increase, too. While the apparent depths of the non-emitting planet transiting single star will remain constant in different bandpasses (neglecting the depth-changing effect caused by stellar limb darkening). Based on the chromatic depth of blending objects, \cite{ODonovan2006} first utilized the multi-color photometry to identify and rule out blending eclipsing binaries in the wide-field transit survey. Later, several \emph{Kepler} follow-up observations eliminated KOIs that displayed chromatic transit depths by ground-based high-precision photometry \citep{Colon2011,Colon2012,Tingley2014,Colon2015} and by parallel \emph{Spitzer} observations \citep{Torres2011,Fressin2012,Desert2015b}. All these successful practices have demonstrated that multi-color photometry is effective in revealing the blends, thereby sparing resources of follow-up observation for other candidates with more scientific interest.

 In this paper, we study how many false positives can be identified by combining an additional reference photometry to \emph{Kepler} bandpass in a \emph{Kepler}-like survey. This double-band photometry observation strategy applies to both simultaneous observation and a follow-up photometry campaign, assuming the reference photometry will accumulate time baseline comparable to \emph{Kepler}'s ($\sim$ 4 year). 
 
 We consider three reference photometric bands: SDSS $z$ \footnote{\url{http://classic.sdss.org/dr7/instruments/imager/index.html}}, 2MASS $K_s$ \citep{Cohen2003}, and TESS \citep{Ricker2014}. As the color-dependent apparent depth is a strong indicator of stellar blends, we use it as the discriminator to identify and rule out the transit-mimicking blending signals, and estimate the overall rule-out fraction of false positives achieved by \emph{Kepler} and chosen reference photometry combinations. In addition, we also evaluate how an extra photometry observation would improve the FPPs of candidates compared to those derived from \emph{Kepler} photometry alone, providing an alternative metric for false positive identification. 

 The false positives considered in this paper are background eclipsing binaries and companion transiting planets, which are most pervasive in transit surveys and may remain undetermined in more sophisticated vetting procedures.
 
 The paper is structured as follows. In Section \ref{sec:fpsim} we define the FPs in consideration and introduce the simulation prescription. In Section \ref{sec:ruleoutfp} we compute the fraction of FPs that can be identified with variable transit depth and study how different stellar properties affect the detectability of depth variations. For those signals which cannot be instantly identified due to insignificant depth variations, we adopt a Bayesian FPP computation framework present the double-band FPP results for typical cases in Section \ref{sec:reducefpp}. We compare these two approaches and discuss some issues in Section \ref{sec:discussion}, and summarize our work in Section \ref{sec:summary}.
 
\section{Simulating the false positive signal on different photometric band}
\label{sec:fpsim}

As the objective of this paper, we investigate chromatic transit depth originating from two typical false positive scenarios: background eclipsing binary (BEB) and companion transiting planet (CTP). For clarity, we define their configuration as follows:

\begin{itemize}
    \item The background eclipsing binary consists of a luminous foreground star (\emph{target star}) and a pair of faint eclipsing binary in the background (\emph{blended companion}).
    \item The companion transiting planet consists of a binary system, with a planet transiting the lower-mass stellar object (\emph{blended companion}), and the signal is largely diluted by the brighter primary star (\emph{target star}).
\end{itemize}

In both cases, the objects that produce the signal are not the \emph{target stars} themselves, but rather are the unidentifiable \emph{blended companion}. We do not consider higher multiplicity systems in our simulation since their occurrence rates are lower than the binary star systems \citep{Raghavan2010}. 

\subsection{Transit depth calculation}
\label{framework}

The observed depth of BEB and CTP on certain wavelength $\lambda$, $\delta_{o,\lambda}$, is only a fraction of the true depth $\delta_{t,\lambda}$,

\begin{equation}
\label{eq:diluteddepth}
    \delta_{o,\lambda} = \frac{F_{b,\lambda}}{F_{b,\lambda}+F_{tar,\lambda}}~\delta_t = d_{\lambda} \cdot \delta_{t,\lambda}
\end{equation}
where $F_{b,\lambda}$ and $F_{tar,\lambda}$ is the fluxes from blending objects and target stars measured on observed wavelength. The ratio between the flux of the blending object and the total flux is denoted as the dilution factor $d_{\lambda}$. If the observed signal comes from a single and isolated system, the dilution factor will be unity. 

The difference between the observed depth in the \emph{Kepler} (KP) and the reference band (RF) can be expressed as

\begin{equation}
\label{eq:depthdifference}
    \Delta \delta_\mathrm{RF-KP} = \delta_\mathrm{t,KP} \cdot d_\mathrm{KP} - \delta_\mathrm{t,RF} \cdot d_\mathrm{RF}
\end{equation}

For a planet transit, the true depth is fixed at the planet-star surface ratio. Therefore the observed depth-difference from CTPs is merely the difference of dilution factors on different photometry. For eclipsing binary, the undiluted primary/secondary eclipse depths will also vary across wavelengths when the binary components have internal temperature difference, and so does the dilution factor, which will further change the depth of the signal.

\subsection{False positive simulation}
\label{sec:fpsim_detail}

To produce the representative population for the \emph{target stars} and the \emph{blended components} defined at the beginning of this section, we generate synthetic stellar populations that are not only physics-based but also consistent with the observational constraints provided by \emph{Kepler} and other surveys. Due to the main purpose of this paper, we need accurate stellar photometry to calculate the apparent depths, which is mainly associated with the stellar masses. We generate these stellar properties with the python package \texttt{isochrones} \citep{2015ascl.soft03010M}, which perform trilinear interpolation in mass-metallicity-age parameter space for Mesa Isochrones and Stellar Tracks (MIST; \cite{Choi2016};\cite{Dotter2016}). \texttt{isochrones} also predicts the stellar photometry using the bolometric correction grid \footnote{\url{http://waps.cfa.harvard.edu/MIST/model_grids.html##bolometric}} provided by MIST. 

\begin{figure}[ht]
    \centering
    \includegraphics[width=\textwidth]{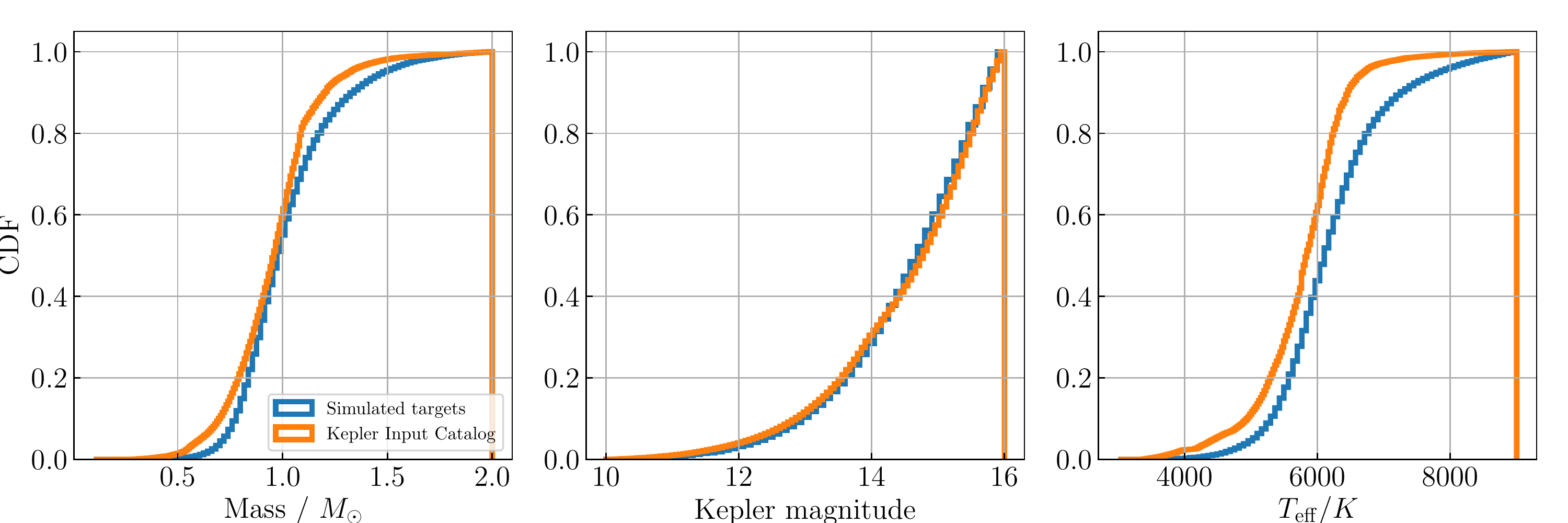}
    \caption{The cumulative density function of target stars' mass, \emph{Kepler} magnitude and effective temperature, compared to the stars in Kepler Input Catalog.}
    \label{Fig1:m_kp_teff_compare}
    \end{figure}

The reference catalog of our \emph{target star} population is Kepler Input Catalog (KIC, \cite{2011AJ....142..112B}), and we use this star population to assign line-of-sight background eclipsing binaries or bounded stellar companions, which will be described in following sections. We list the simulation parameters of \emph{target star} population in Table \ref{tab:target_star_prop}. We compare the distributions of mass, apparent \emph{Kepler} magnitude, and effective temperature of the \emph{target star} population, along with those of KIC stars, in Figure \ref{Fig1:m_kp_teff_compare}. The agreement between KIC and our \emph{target star} population is satisfying, though our simulation contains more hotter stars. We will adress this difference and its impact in Section \ref{sec:diff_prop_kic}.

\begin{table}[ht]
    \begin{center}
    \caption{Input parameters of \emph{target star} population}
    \label{tab:target_star_prop}
    \begin{tabular}{ll}
      \hline\noalign{\smallskip}
        Initial mass function (IMF)      & Chabrier IMF       \\
        {[}Fe/H{]}                       & Double-Gaussian in \citep{Casagrande2011}     \\
        Age        & Uniform(0,10) Gyr       \\
        Av$^a$         & 0.1-1.6 mag $~\cdot \mathrm{kpc}^{-1}$    \\
        Distance   & $dN \sim d^2,~d \in [0,1800]$ pc \\
        Magnitude limit (\emph{Kepler}) & 16 \\
        
      \noalign{\smallskip}\hline
    \end{tabular}
    
    \footnotesize{$a$. Drawn from \url{https://ned.ipac.caltech.edu/extinction_calculator}, depending on the observing location.}
    \end{center}
    \end{table}

To infer the stellar properties of background stars and their spatial gradients in background fields, we use TRILEGAL to simulate background stars \citep{Girardi2005}. TRILEGAL is a web-based Milky Way star population simulator that considers different galactic components. The request can be called by setting the coordinate of desired fields and galactic structures (we use defaults of structure parameters). We use the stellar mass and distance provided by TRILEGAL to generate eclipsing binaries for BEB simulation. 

We provide detailed descriptions about how we construct the appropriate BEBs and CTPs in following Section \ref{sec:bebsim} and Section \ref{sec:ctpsim}.

\subsubsection{Simulating background eclipsing binaries}
\label{sec:bebsim}

    We downloaded one-square-degree background stars in the centers at each of the 21 modules in \emph{Kepler} field of view via TRILEGAL. For the sake of consistency in deriving stellar properties as we generate \emph{target stats} populations, we extract the returned parameters, e.g., initial mass, metallicity, age, distance, and extinction factor, from TRILEGAL and pass them to \texttt{ischrones} to obtain the multi-band photometry and radius of the background primaries. Next, according to the flat mass-ratio and the binary fraction as the function of primary's spectral type reported in \cite{Raghavan2010}, we generate secondary stars to selected primaries.
    
    The eclipse of a binary system only occurs when the impact parameter meets the criterion \citep{Winn2010}:
    
    \begin{equation}
     b = \frac{a~\cos i}{R_1} \frac{1-e^2\cos \omega}{1 \pm e \cos\omega} < 1 + \frac{R_2}{R_1}
    \end{equation}
    where ``$ + $'' stands for the presence of primary eclipse and ``$-$'' for secondary eclipse, $b$ for the impact parameter of the eclipse, $a$ for the semi-major axis, $R_1,~R_2$ for the radius of primary and secondary star, $e$ for eccentricity, $i$ for inclination and $\omega$ for argument of periastron. We assume the orbital elements of each binary pair follow such distributions:
    
    \begin{itemize}
        \item Period and eccentricity distribution from Multiple Star Catalog \citep{Tokovinin2018}, and truncate the period up to 300 days.
        \item Uniform distribution between $0 \sim 360^\circ$ for $\omega$ and a cosine distribution for orbital inclination.
        \item Reject the systems that are in each other's Roche Lobe (Equation 2 in \cite{1983ApJ...268..368E}). 
    \end{itemize}
    
     For each background eclipsing binary, we assign a foreground star from the \emph{target star} population, and then we calculate the diluted primary/secondary eclipse according to Equation \ref{eq:diluteddepth}. Next, we calculate the diluted eclipse's signal-to-noise ratio (SNR) according to
     
     \begin{equation}
     \label{eq:snr}
         SNR_\mathrm{pri/sec,KP} = \frac{\delta_\mathrm{pri/sec,KP}}{\sigma_\mathrm{eff, KP}},~\sigma_\mathrm{eff, KP}= \sigma_\mathrm{6hr-CDPP}~\sqrt{\frac{T_\mathrm{pri/sec}}{6~hr}~\frac{P}{4~yr}}
     \end{equation}
     where $\delta_\mathrm{pri/sec,KP}$ is the observed primary/secondary depth on \emph{Kepler} band, $\sigma_\mathrm{eff, KP}$ is the effective photometric precision, which is based on the target star's 6-hour Combined Differential Photometric Precision (CDPP) inferred from, eclipse duration $T_\mathrm{pri/sec}$ and the orbital period $P$, and decide whether they are detectable on \emph{Kepler} band. To model the signal detection condition of \emph{Kepler}, we reject BEBs whose both of the eclipses' SNR$_\mathrm{KP}<7$ \citep{2010ApJ...713L..87J}, which is the nominal detection threshold of \emph{Kepler} pipeline to avoid misidentification due to statistical fluctuation in the ensemble of all target stars. 
     
     Central for a BEB to mimic planet-transit is it must not show distinguishable difference between primary and secondary eclipse. Based on the calculated SNR in \emph{Kepler} bandpass, we classify the BEBs that have passed the secondary eclipse test into two categories:
    
    \begin{itemize}
        \item Type 1: the primary and secondary eclipse are both detectable, but are indistinguishable in depths (within $3\sigma$), which are taken as candidates with periods of half of the actual eclipse period, and the number of detected `transits' will be doubled. Thus, the detection threshold of this type of BEB will be $\frac{7}{\sqrt{2}}\sim 5$.
        \item Type 2: one eclipse is `missing' in the observation, which may result from high-eccentric orbits, or the primary/secondary eclipse is diluted to be undetectable (SNR$_\mathrm{pri/sec} < 3$) by the foreground star while the other eclipse is still detectable.
    \end{itemize}
    
    The two groups of BEBs are well separated and distinctive on the SNR$_\mathrm{KP}$ diagram of primary/secondary eclipse. An example of the SNR$_\mathrm{KP}$ diagram is shown in Figure \ref{fig:beb_prop}. Type 1 BEBs lie in the diagonal of the diagram, with a median orbital eccentricity of 0.01 and a mass ratio of 0.89. Type 2 BEBs cluster along the `two wings' symmetric to the diagonal, with a median orbital eccentricity of 0.39 and mass-ratio of 0.52.

    \subsubsection{Simulating companion transiting planets}
    \label{sec:ctpsim}
    
    The CTP model consists of a planet transiting the secondary star of the binary system. To construct the binary components, we select stars from the \emph{target population} as the primary stars, in which we account for the reported binary fraction's dependence on primary mass. Then, we assign secondary companions to the primaries with a uniform mass ratio between 0.1 and 1 \citep{Morton2011,Raghavan2010}. We assume the binary members are co-located and coeval, that is, the secondary stars have the same metallicity, age and distance as primaries. 
    
    Next, we assign a planet to each secondary star. The planets are parameterized by their radii and orbital periods. Orbital periods are randomly selected from KOI lists \citep{Batalha2013}. The assignment of planet radii, however, is not so straightforward. We will address this issue in detail in Section \ref{sec:ctp}.

\subsection{Reference bands and noise model}
    \label{sec:noisemodel}

    In our work, we have chosen three photometric bands to be the reference band of \emph{Kepler}, and each of them has a distinctive effective wavelength and coverage: SDSS $z$ (800-900 nm), 2MASS $K_s$ (1.9-2.3$ \mu$m), and TESS (600-900 nm) band. SDSS $z$ band is a narrow-band filter commonly used in ground-based photometry. TESS band is wide-band photometry is employed by the spacecraft TESS especially designed for transit photometry. 2MASS $K_s$ is a near-infrared filter band and may represent future space-based infrared photometry. Our preference for longer-wavelength photometry partly stems from the assumption about the considered false positives: we assume the \emph{Kepler} targets are often blended with a lower-mass object, which is a lot redder in colors (and interstellar extinction will further redden the color of background binaries in BEB). Hence, a combination of photometry bands with concentration at visible light (such as \emph{Kepler} band) and infrared (and even near-infrared) would properly exploit and reveal the color-difference in BEBs and CTPs. The aforementioned assumption is reasonable for most of the stubborn cases of FPs that are of relatively small sizes and low brightness. Target stars blended with higher mass (and thus brighter) objects may have already been compiled in the KIC catalog or identified by other techniques, such as centroid analysis \citep{Bryson2013} or ephemeris matching \citep{Coughlin2014}.
    
\begin{figure}[ht]
        \centering
        \includegraphics[width=0.5\textwidth]{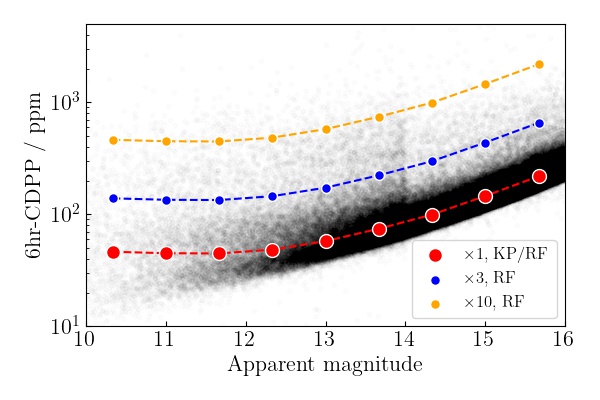}
        \caption{The adopted photometric precision relation to evaluate the observed depth uncertainties for our simulated false positives (red, blue and yellow dashed curves). The photometric uncertainty on \emph{Kepler} band is represented by the red curve, which is the median 6-hr CDPPs (black dots) in 9 bins among all stars in Kepler Input Catalog with $m_\mathrm{KP}$ between 10 and 16. The photometric precision on reference bands ($z,~K_s,~\mathrm{TESS}$) is evaluated by the apparent reference magnitude $m_{RF}$ according to the three example precision curves, which are the scaled version of \emph{Kepler} precision with factors of one, three, and ten.}
        \label{fig:noisemodel}
    \end{figure}

    To quantify the uncertainties in transit depth measurements and the significance of the depth variations in different bandpasses, it is necessary to assign photometric precision to each simulated FP instance. The Combined Differential Photometric Precision (CDPP) measured in \emph{Kepler} stars provides the best template for our purpose. Here we assume the CDPP is only correlated to the \emph{Kepler} magnitudes $m_\mathrm{KP}$. To derive an empirical relation, we take median 6-hour CDPP evaluated in 1-magnitude bins between $m_\mathrm{KP}=9$ and 16. 
    We assign \emph{Kepler} noise to each FP instance in terms of its apparent $m_\mathrm{KP}$ by linearly interpolating the CDPP-$m_\mathrm{KP}$ relation explained above. For the reference bands, we multiply \emph{Kepler} 's CDPP-$m_\mathrm{KP}$ curve by factors of one, three, and ten, interpolating the scaled noise-$m_\mathrm{RF}$ relation based on the apparent $m_\mathrm{RF}$. By adopting different noise templates can we provide a fiducial estimate for the photometry precision for the reference band: when $\sigma_\mathrm{RF}=\sigma_\mathrm{KP}$, the reference photometry attain the best space-based photometry precision so far, as have been examplified by \emph{Kepler}. If $\sigma_\mathrm{RF}=3\sigma_\mathrm{KP}$, this is another typical precision for space-based telescope with smaller aperture, like TESS (30cm). In the case of $\sigma_\mathrm{RF}=10\sigma_\mathrm{KP}$, this is where the typical ground-based photometry can achieve. Examples of \emph{Kepler} and reference band noise model are presented in Figure \ref{fig:noisemodel}.
    
\section{Ruling out false positives with depth-difference}
\label{sec:ruleoutfp}

\subsection{Background eclipsing binaries}
\label{sec:beb}
	\subsubsection{Identify BEBs with two photometric band}

    This section demonstrates that the \emph{Kepler}'s candidates mimicked by background eclipsing binaries can be identified and excluded by joining additional reference photometry. In the previous section, we define those BEBs have passed the secondary eclipse test on \emph{Kepler} band will be indistinguishable from genuine planet transits, i.e., on the \emph{Kepler} band, the primary's and secondary eclipse have similar depths, or only primary/secondary eclipse is detectable. Since a joint photometry is observing with \emph{Kepler}, we may also check whether these BEBs can also pass the secondary eclipse test on the reference band. We then identify the BEBs for which the SNR of primary-secondary eclipse difference on reference band, SNR$_\mathrm{pri-sec,RF}$, is larger than 5. This quantity is expressed in Equation \ref{eq:beb_prisec_diff}. 
    
    \begin{equation}
    \label{eq:beb_prisec_diff}
        SNR_\mathrm{pri-sec,RF} = \frac{|\delta_\mathrm{pri,RF}-\delta_\mathrm{sec,RF}|}{\sqrt{\sigma_\mathrm{pri,RF}^2+\sigma_\mathrm{sec,RF}^2}} > 5
    \end{equation}
	where $\delta_\mathrm{pri,RF}$ and $\delta_\mathrm{sec,RF}$ is the diluted primary and secondary depth of BEB, and $\sigma_\mathrm{pri,RF},~\sigma_\mathrm{sec,RF}$ is the effective noise of primary/secondary eclipses. Some BEBs may pass the secondary eclipse test on both photometry bands. To eliminate those stubborn cases, we proceed to compare the apparent depth on two photometry bands. To estimate the number of simulated BEBs that show detectable depth variations between \emph{Kepler} and the reference band, we calculate the SNR of depth-difference, SNR$_\mathrm{RF-KP}$, and identify those BEBs whose SNR$_\mathrm{RF-KP}$ is larger than 5. This quantity can be expressed as:
    
    \begin{equation}
    \label{detectabledepthdifference}
        SNR_\mathrm{RF-KP} =  \frac{|\delta_\mathrm{RF}-\delta_\mathrm{KP}|}{\sqrt{\sigma_\mathrm{RF}^2+\sigma_\mathrm{KP}^2}} > 5
    \end{equation}
    where $\delta_\mathrm{RF},~\delta_\mathrm{KP}$ are the apparent transit depth observed in reference band and \emph{Kepler}.
    
\begin{figure}
        \centering
        \begin{minipage}[t]{0.475\linewidth}
        \centering
         \includegraphics[width=1\textwidth]{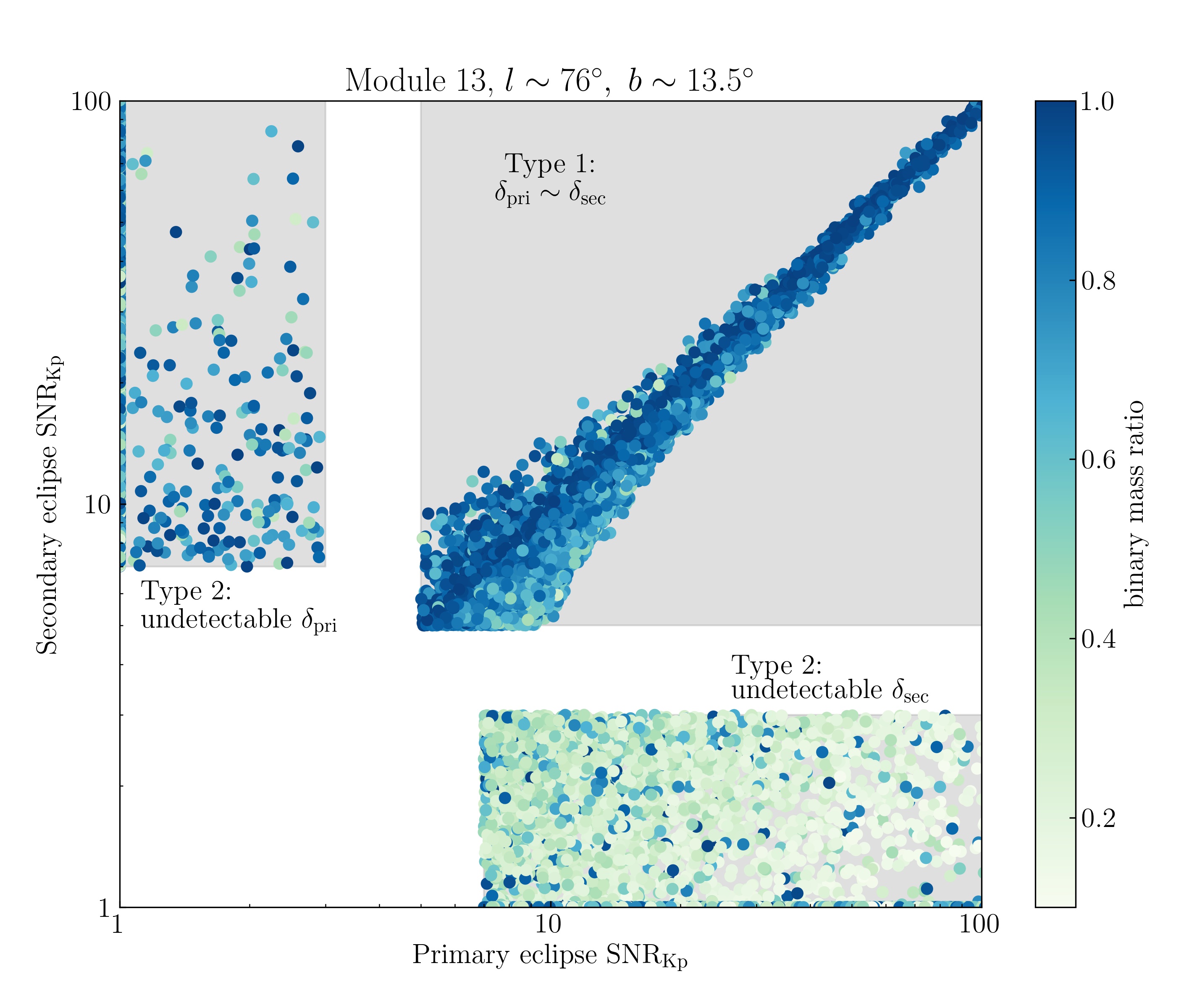}
        \end{minipage}%
        \begin{minipage}[t]{0.4\textwidth}
        \centering
         \includegraphics[width=1\textwidth]{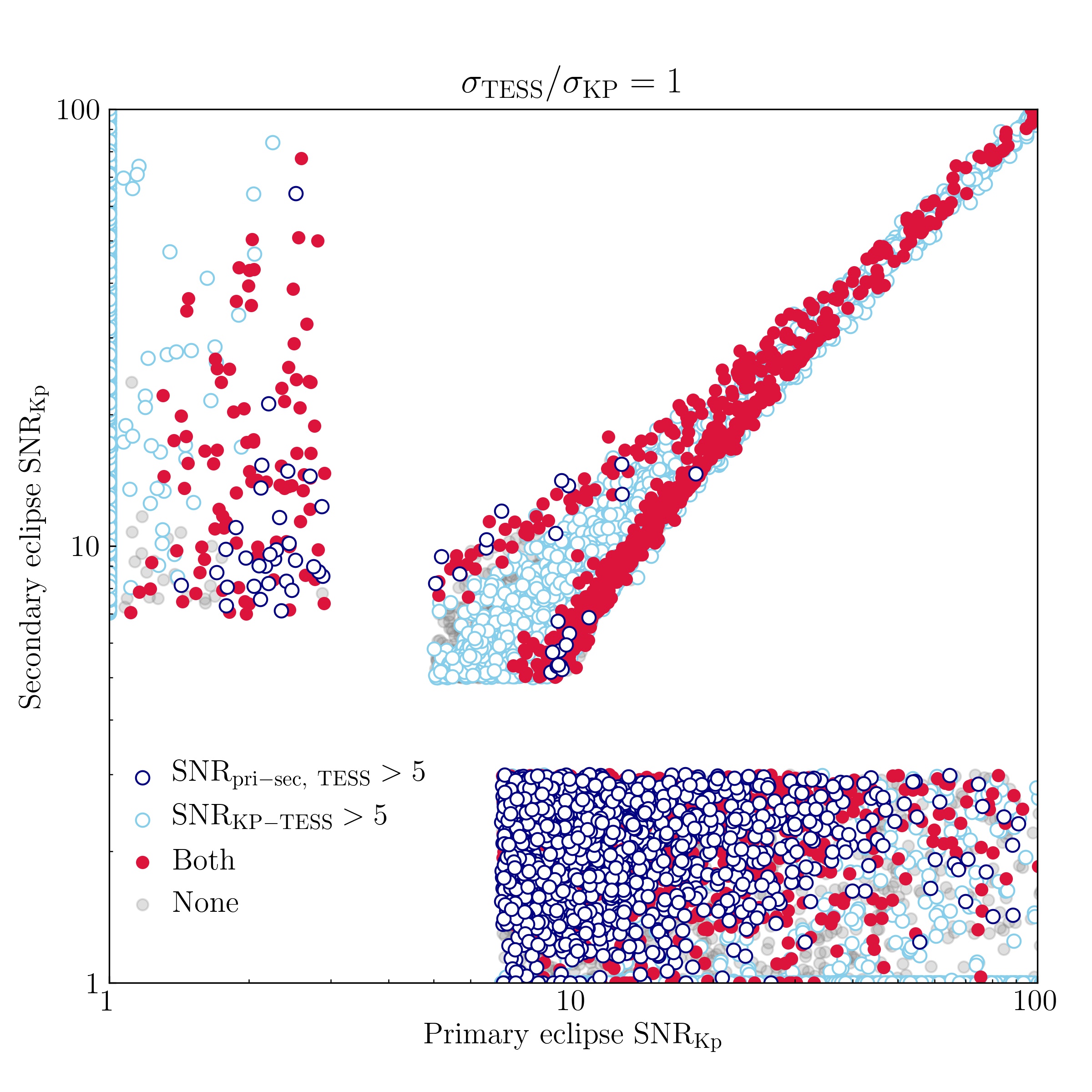}
        \end{minipage}%
        \caption{\emph{Left}: The SNR$_\mathrm{KP}$ of the primary and secondary eclipse of simulated background eclipsing binaries (BEBs), color-coded by the mass-ratio $q$ of background binary. Type 1 BEBs has undistinguishable primary and secondary eclipses in \emph{Kepler} photometry. They lie in the diagonal and consist of mainly equal mass ($q\sim0.88$) and circular-orbit ($e\sim0.01$) background binaries . Type 2 BEBs have one undetectable eclipse. They cluster in the upper left and bottom right corner of the plot, and they are mainly composed of low mass-ratio ($q\sim0.43$) and eccentric-orbit ($e\sim0.26$) binaries; \emph{Right}: the same as the left figure, but the BEBs are marked in terms of the detectable double-band photometry feature. Light blue circles are the BEBs that only show distinguishable depth variations (SNR$_\mathrm{RF-KP}>5$), which are mainly Type 1 BEBs and low $q$ Type 2 BEBs; deep blue circles are those BEBs that can only be revealed by their identifiable primary-secondary eclipse difference on the reference band, which are mainly binaries with intermediate $q$; red filled circles are the BEBs that both of the signatures, and the grey circles show neither depth variation nor primary-secondary difference. The instances whose SNR$_\mathrm{KP}$ of primary or secondary eclipse is below 1 are plotted on the edge of the figure, for visualization. The reference band used to plot this figure is TESS, assigned with a noise estimate identical with \emph{Kepler}'s. }
        \label{fig:beb_prop}
    \end{figure}

    With either of these two features observed, one can confidently identify the blending nature of the transit signals. We present an example of the SNR of primary-secondary difference and double-band depth variations for the BEBs using \emph{Kepler}-TESS pair photometry, which is displayed in the left panel of Figure \ref{fig:beb_prop}, and we mark the BEBs if they show any of the two characteristics in the right panel. The majority of Type 1 BEB can only show double-band depth variations since most of them contain  equal-mass EBs so that the primary and secondary eclipse will remain nearly identical across all photometry. The detectable primary-secondary difference is preferentially displayed by BEBs with intermediate mass-ratio, primarily Type 2 BEB.
    
\begin{figure}
        \centering
        \includegraphics[width=1\textwidth]{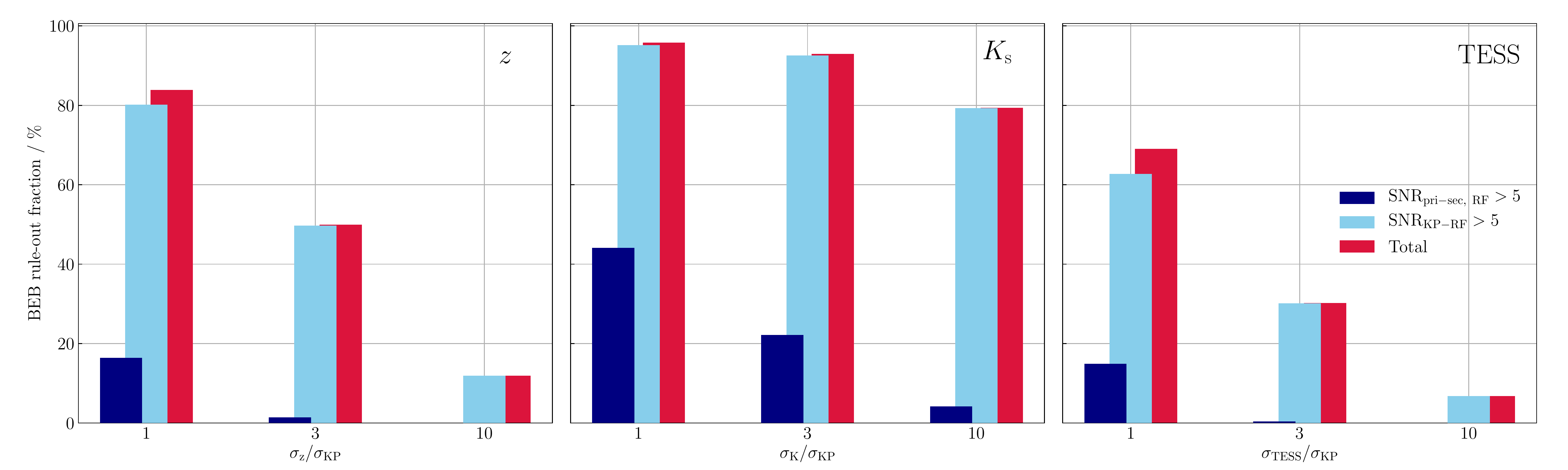}
        \caption{The fraction of BEBs showing detectable depth variations or detectable primary-secondary difference or both, as a function of the photometric precision on reference band. }
        \label{fig:beb_ruleout}
    \end{figure}

    Figure \ref{fig:beb_ruleout} compares these two identification methods at different reference bands and precision. Generally, more BEBs will show identifiable depth variations rather than primary-secondary difference. On the $K_s$ band, most of the BEBs will show both of the features, and this fraction only differs between 80-95\% under the considered reference photometry precision. In contrast, using $z$ and TESS band can at most identify 83\% and 69\% of the BEBs, respectively. And the fractional number will drop sharply to only around 12\% and 6.8\% when $\sigma_\mathrm{RF}/\sigma_\mathrm{KP}=10$. Moreover, the fraction of BEBs identified via primary-secondary difference is nearly negligible under relatively large noise ($\sigma_\mathrm{RF}/\sigma_\mathrm{KP}=3,10$) for $z$ and TESS band.

	\subsubsection{Factors affecting the the amplitude of depth variations}
        
\begin{figure}[ht]
        \centering
        \includegraphics[width=1.0\textwidth]{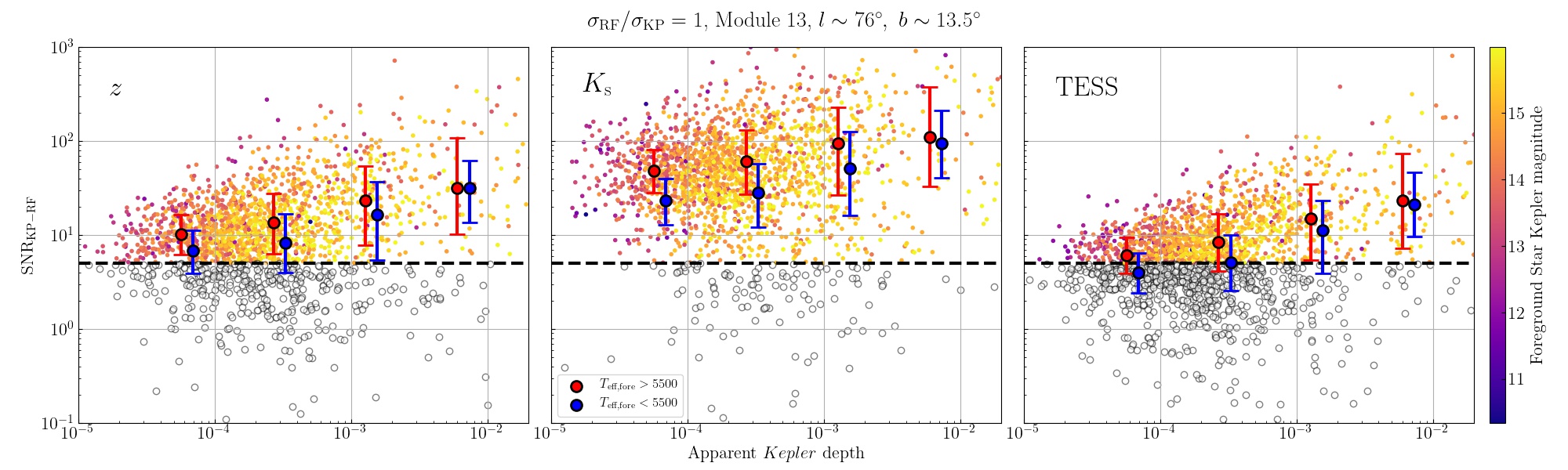}
        \caption{BEBs' SNR$_\mathrm{RF-KP}$ as a function of the apparent \emph{Kepler} depth, color-coded by the \emph{Kepler} magnitude of foreground star. The black dashed line separate the detectable depth-difference (SNR$_\mathrm{RF-KP}>5$) from those undetectable. BEBs with fainter foreground stars tend to produce deeper signals with larger SNR$_\mathrm{RF-KP}$. Red and blue points represent the median SNR value of BEB subsets with hot and cold foreground star (respectively) in each apparent depth bin. Error bars correspond to 25th and 75th percentile. Hotter foreground stars may have larger color-difference relative to the background binaries. Therefore they can be more easily distinguished by double-band photometry.}
        \label{fig:beb_SNR_vs_kpdepth}
    \end{figure}

    According to Figure \ref{fig:beb_ruleout}, most BEBs in our simulation show significant double-band depth difference rather than primary/secondary difference. We may examine how the amplitude of depth-difference is correlated with BEB system properties. We present the SNR$_\mathrm{RF-KP}$ of all BEBs in Figure \ref{fig:beb_SNR_vs_kpdepth}, along with the apparent depths and \emph{Kepler} magnitude exhibited by the systems. We find that BEBs preferentially produce and mimick Earths and small-Neptune candidates. This pattern agrees well with the simulation of \cite{Fressin2013}. Besides, BEBs with large apparent \emph{Kepler} depth tend to show large SNR$_\mathrm{RF-KP}$, even though they are produced by faint foreground target stars. However, even for small planet candidates, at least half of them can display detectable depth variations, assuming the reference bands can achieve a \emph{Kepler}-like precision.
    
    The color-difference between different stellar components in BEB configurations will also influence the amplitude and detectability of the wavelength-dependent depth feature: (i). the color-difference (temperature difference) between the foreground target star and background system, in which foreground stars' dilution level will differ between bandpasses and (ii). the internal color-difference (or equivalently the effective temperature difference) between the background binary will result in a chromatic eclipse depth. With such a view, we may proceed to explore the color-difference between which stellar components would impact SNR$_\mathrm{RF-KP}$ more significantly.
    
\begin{figure}[ht]
        \centering
        \includegraphics[width=1.0\textwidth]{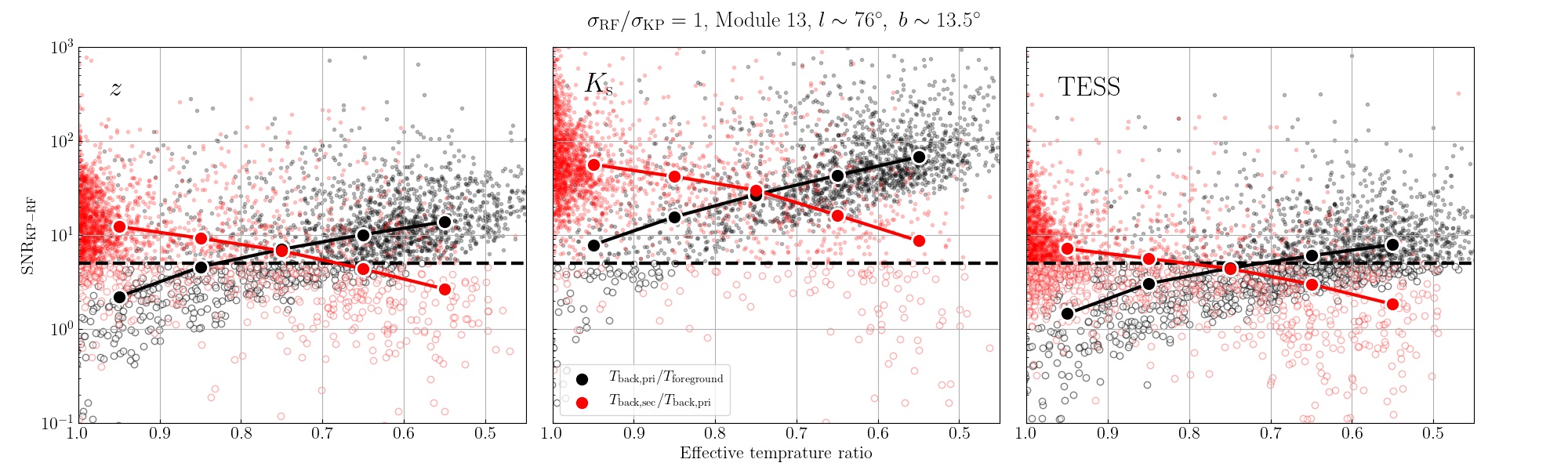}
        \caption{BEBs' SNR$_\mathrm{RF-KP}$ plotted against the effective temperature ratio between different stellar components. Black dots are the temperature ratio between background primaries and foreground stars; red dots are the temperature ratio between background secondaries and primaries. The median SNR$_\mathrm{RF-KP}$ calculated  in each $T_\mathrm{eff}$-ratio bin is represented by big dots. The figure suggests that the larger color-difference between foreground and background stars will enhance the detectability of depth variation. However, the color-difference between background binary's negatively impacts the SNR$_\mathrm{RF-KP}$. The opposite trend results from the decreasing eclipse depth when the secondary star is much smaller and fainter in low $T_\mathrm{eff}$-ratio region.}
        \label{fig:tratio_SNR12}
    \end{figure}

    To better illustrate this, we plot BEBs' SNR$_\mathrm{RF-KP}$ against the temperature-ratio between background binary components and foreground-background stars separately in Figure \ref{fig:tratio_SNR12}. The effect of (i) is evidenced by the positive trend between the SNR$_\mathrm{RF-KP}$ and background-foreground star temperature ratio (black dots) presented in Figure \ref{fig:tratio_SNR12}. The effect of (ii), on the contrary, show a negative correlation (red dots) with SNR$_\mathrm{RF-KP}$, as shown in Figure \ref{fig:tratio_SNR12}. This is because the effective temperature not only correlates to the color of star, but also with star's size and luminosity. As the secondary star gets a lot cooler than the primary, it will become smaller and less luminous, therefore the largest possible  decline in brightness during the eclipse will drop, too. The decrease of the true eclipse depth will reduce the magnitude of apparent diluted depth and the amplitude of depth variations. Thus, BEBs that consist of background binary members with distinct spectral types tend to show minor amplitude in depth variations, even if the internal color-difference is large.
    
    \subsubsection{Ruling out BEBs with different apparent depths}
    
\begin{figure}[ht]
        \centering
        \includegraphics[width=0.9\textwidth]{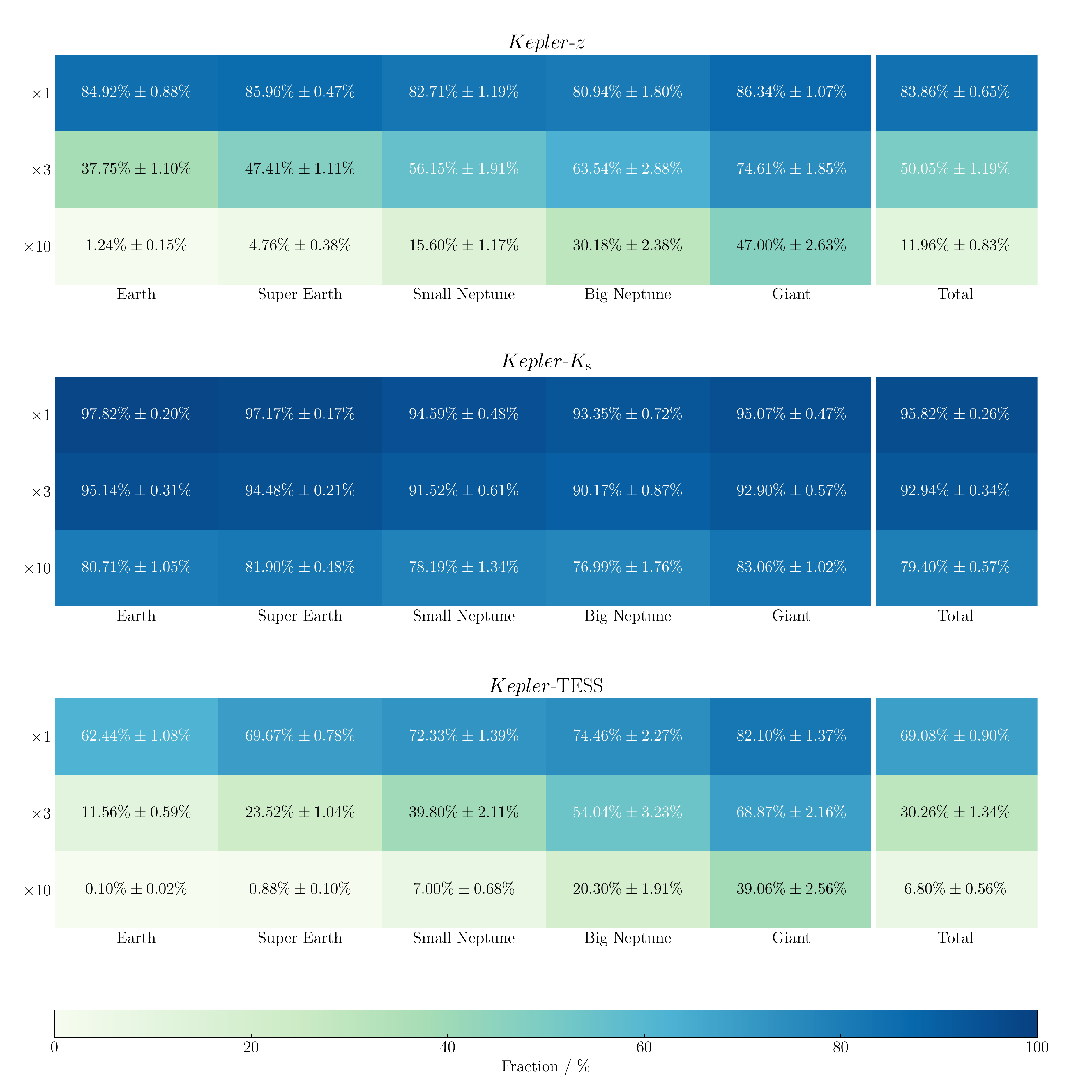}
        \caption{This figure sums up the BEBs' rule-out fraction, classified by different apparent planet radii on \emph{Kepler} band. The number on left column denotes the adopted photometric precision on the reference band. The numbers in each cell are the mean value of rule-out rate, averaged over the simulation from 21 fields centered on \emph{Kepler}'s CCD modules. The uncertainty represents 95\% confidence interval. Our planet radii classification is the same as that in \cite{Fressin2013}. $K_s$ band can indiscriminately identify the BEBs mimicking all sizes of planet radii. TESS's and $z$ band's successful identification prefers giant planet candidates mimicked by BEBs, though $z$ is generally better than TESS.}
        \label{fig:BEB_plradius_ruleout}
    \end{figure}

    We learn that different types of BEBs can be identified through different approaches, and the detectability is correlated to the stellar properties in the system. We may predict how much BEBs in \emph{Kepler} can be identified overall. The mass distribution of background star have gradient along the galactic latitude covered by \emph{Kepler} Field of View (FoV); thus such mass variance may affect the BEB's rule-out fraction in different modules in the entire \emph{Kepler} field. To estimate the overall rule-out fraction, we simulate BEB population in the centers of 21 modules of \emph{Kepler} FoV via TRILEGAL, while the pool of foreground stars are the same as introduced in Section \ref{sec:bebsim}. We calculate the rule-out fraction and in each field. We present the average fraction in Figure \ref{fig:BEB_plradius_ruleout}, sorted by different reference bands, the size of planet candidate they mimic (based on apparent \emph{Kepler} depth), and the assumed reference photometry precision. 
    
    The \emph{Kepler}-$K_s$ photometry pair can efficiently rule out all BEBs with any depths. Even if the depth comparison is done with the largest noise on the reference band, the rule-out rate can still attain nearly 80\% for our BEB samples. 

    The rule-out fractions achieved by \emph{Kepler}-TESS photometry pair strongly depend the apparent depths that the BEBs display. Although 80\% of BEBs that mimics giant planet candidates can be effectively ruled out if TESS can achieve a \emph{Kepler}-like precision, only half of the BEBs with Earth-size depth can be identified. Since most BEBs' apparent depth is shallow, the overall rule-out fraction is only around 69\%, still below that of the $Kepler-K_s$ pair with the largest noise. Moreover, the exact level of $\sigma_\mathrm{TESS}$ will significantly affect the rule out rate of Earth-like BEBs, too. When $\sigma_\mathrm{TESS}$ is three times of \emph{Kepler} precision, the rule-out rate for Earth-like BEBs drops to 12\%. In the worst precision, nearly none of Earth-like BEBs can be identified, and only those mimicking giant planet candidates are identifiable. The performance of \emph{Kepler}-$z$ photometry pair better than TESS. If $z$ band can achieve a \emph{Kepler}-like precision, it can rule out 80\% of the BEBs with any apparent depths. However, shallow-depth BEBs get harder to identify as the noise on $z$ band gets larger, and thus the total rule-out rate decreases.

\subsection{Companion transiting planets}
\label{sec:ctp}

    The search of planet transit is often carried out without knowing whether the target stars are single are not. Unrecognized binarity of the target star will bring uncertainty to the planet radius inferred from the observed signal. If the planet is transiting the primary star, the inferred planet radius is underrated by $\sqrt{2}$ at most (in case of equal-mass binary) and is thus not seriously affected. Also, there may well be a chance that this is actually a large planet transiting an unseen stellar companion, and the transit depth is diluted to appear much smaller due to primary's flux contamination. In this case, the radius correction $X_R$, the ratio of the planet's true radius to apparent radius, can be up to six \citep{Ciardi2015}. Multi-color photometry can help identify the binarity of the system by identifying the depth variation in transit light curves.

    Part of the objective of this paper is to estimate the exclusion rate of \emph{Kepler}-detected false positives via double-band photometry. However, deriving the exclusion rate in the case of companion transiting planets is not so straightforward as background eclipsing binaries. This is because currently, we lack the specific radius distribution of planets that are transiting the secondary star in a binary system, $P^\mathrm{CTP} (R_p)$. Simply extrapolating the radius distribution from other work based on completely different contexts would undermine the reliability of our result. Thus, we refrain from employing any functional forms of the planet radius distributions in our simulation, nor will we derive a definitive value of the exclusion rate. Instead, in this section, we aim to derive the specific exclusion rate as the function of the true planet radius, i.e., the probability that a CTP system can show detectable depth variations if the hosting planet has a radius of $R_p$. These formalisms can accommodate any forms of $P^\mathrm{CTP} (R_p)$ to derive the overall exclusion rate. We will first discuss how different spectral-type binary would influence the specific exclusion rate in Section \ref{sec:ctp_spec_types}, then we present the final formalism in Section \ref{sec:spec_exclusion_rate}. 
    
    \subsubsection{Depth variation detection probabilty for CTPs with different spectral types}
    \label{sec:ctp_spec_types}

    According to Equation \ref{eq:depthdifference}, the amplitude of depth variation of CTP model comes from the color-difference between the binary members and the radius ratio between the transiting planet and secondary star. In this regard, we may expect the amplitude of CTP's depth variation is correlated to the spectral types of binary members and the true radius of the planet. However, the detectability of depth variations not only depends on the amplitude of variations but also on the planet orbital periods and apparent \emph{Kepler} magnitude, which are needed to calculate the effective photometry precision.

\begin{figure}[ht]
        \centering
        \includegraphics[width=1\textwidth]{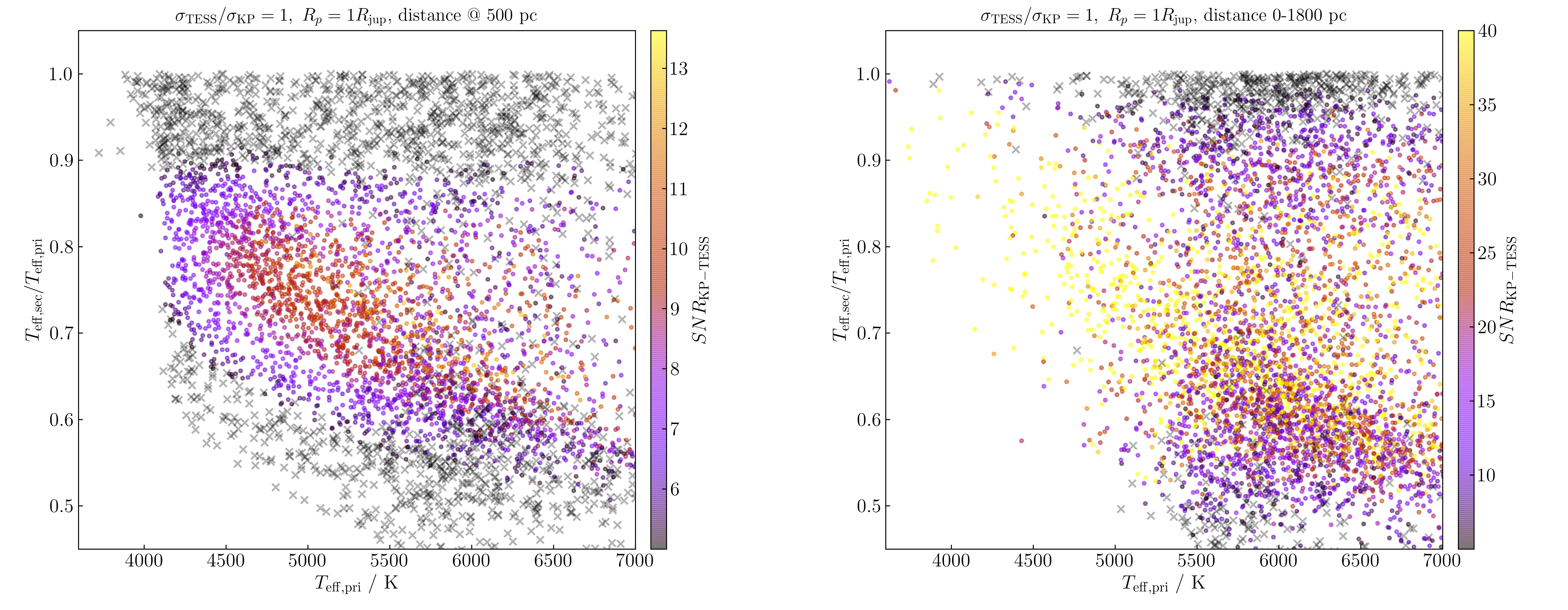}
        \caption{\emph{Left}: an example of the SNR$_\mathrm{TESS-KP}$ distribution of equal-distance ($\sim$ 500 pc) CTP simulations plotted against the primary star effective temperature $T_\mathrm{eff,~pri}$ versus the primary-secondary temperature ratio $T_\mathrm{eff,~sec}/T_\mathrm{eff,~pri}$. CTPs showing undetectable depth variations (black crosses) are mainly binary systems having identical primary and secondary stars. Therefore there is little color-difference in between. The other part of the diagram shows a U-shape distribution at a fixed value of SNR$_\mathrm{TESS-KP}$, which is the result of primary dilution effect and stellar color-difference; \emph{Right}: the same as the left figure, but the CTPs are spatially distributed far out to 1800 pc. The spatial distribution will randomly decrease the photometric noise of the system}
        \label{fig:ctp_snr}
    \end{figure}

    To start with, we simulate a simplified population of CTPs, which all locate at the same distance (500 pc) and uniformly have a Jupiter-size planet. We calculate the SNR$_\mathrm{TESS-KP}$ of one-transit signal, so that the SNR$_\mathrm{TESS-KP}$ is mostly determined by binary's spectral type composition. In the left panel of Figure \ref{fig:ctp_snr}, we present these CTPs' SNR$_\mathrm{TESS-KP}$ as well as the effective temperatures of binary components. It can be seen that CTP instances with similar SNR$_\mathrm{TESS-KP}$ form a distinctive U-shape pattern, which suggests these CTPs, despite that they are made up of different spectral types of stars, have a similar amplitude of depth variation. For a given primary star, the SNR$_\mathrm{TESS-KP}$ increases as the $T_\mathrm{sec}/T_\mathrm{pri}$ decreases from one, in which the binary components have more discrepant color and the amplitude of depth variation subsequently increases, too. Then SNR$_\mathrm{TESS-KP}$ decreases as temperature-ratio keeps dropping after some point. When the primary star is much more massive than the secondary star, its flux contamination is so heavy that the apparent transit depth is attenuated on both photometry band, and the depth-difference will be small. The primary's dilution effect on SNR$_\mathrm{TESS-KP}$ is most extreme at the bottom of the figure, where the binaries are mainly an M dwarf bound to an F/G dwarf, rendering the depth variations to be almost undetectable. 

    The U-shape pattern no longer holds if we add in distance and planet orbital distributions to CTPs (the right panel of Figure \ref{fig:ctp_snr}), which introduce fluctuations to the effective precision of individual CTP system. However, the absolute amplitude of CTP depth variation is determined by the system's spectral type composition. Therefore we may expect there are certain populations of CTPs that are most amenable to display identifiable depth variation. 
    
    To evaluate the probability that the depth-difference signal arising from a binary system with certain spectral types can attain sufficient SNR$_\mathrm{RF-KP}$ in a \emph{Kepler}-like survey. We redo the simulation in the right panel of Figure \ref{fig:ctp_snr}; in each trial we change the binaries' distances and planets' orbital periods. We calculate the fraction of systems whose SNR$_\mathrm{RF-KP}$ is larger than 5 in gridded $T_\mathrm{pri}-T_\mathrm{sec}/T_\mathrm{pri}$ space.  Finally we average the value over ten trials.  
    
\begin{figure}[ht]
        \centering
        \includegraphics[width=1\textwidth]{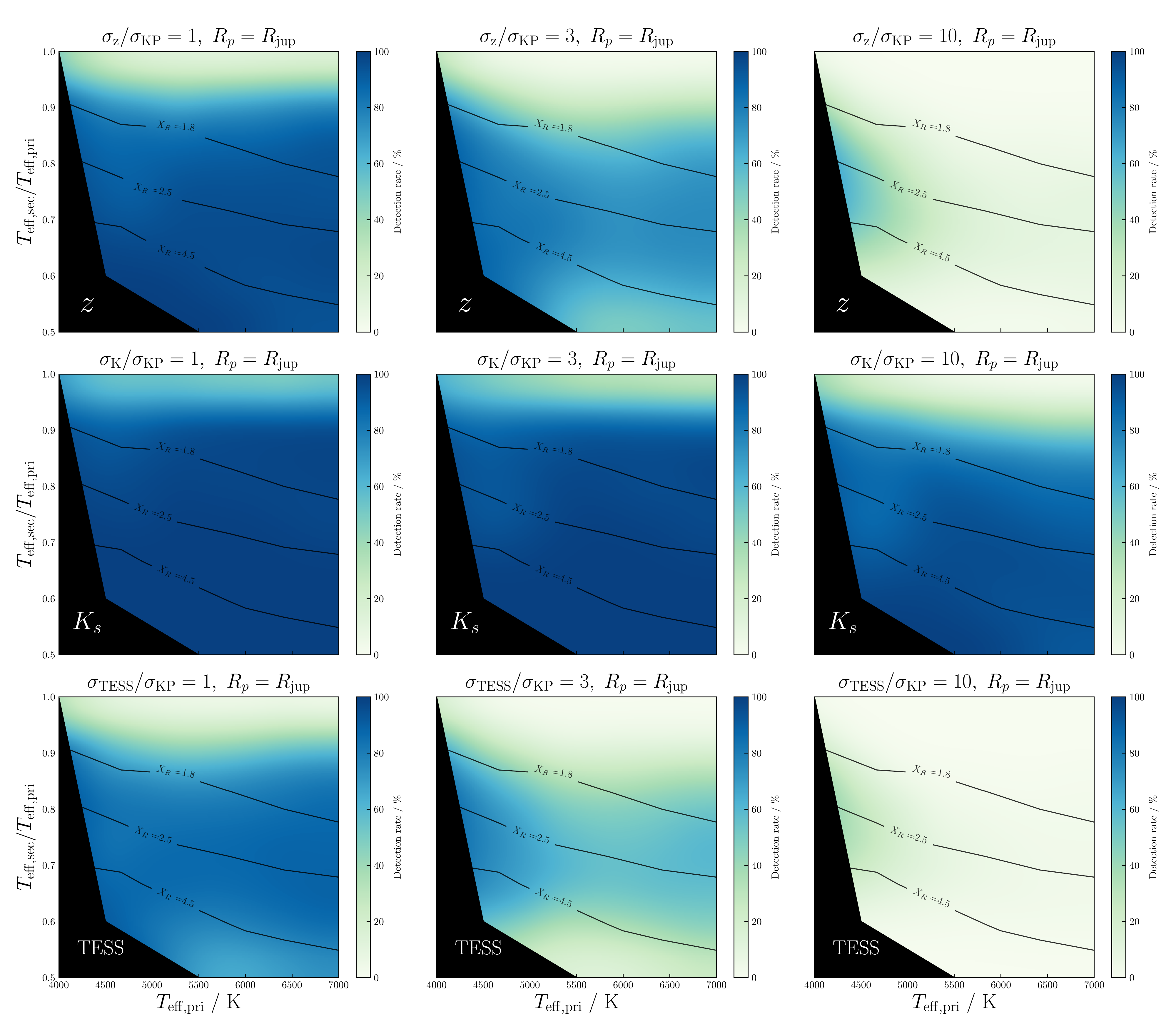}
        \caption{The probability that a CTP system show detectable depth variations, in terms of the effective temperature of primaries and secondaries. The probability is evaluated in a grid defined in $T_\mathrm{pri}-T_\mathrm{sec}/T_\mathrm{pri}$ space, averaged over 10 trials for each reference band and noise. Blocked area is the region where the mass of secondary star is below 0.1$M_\odot$. Contours are the planet radius correction factor defined as the same in \cite{Ziegler2016}.}
        \label{fig:ctpdetectionrate}
\end{figure}

    We present the depth variation detection probabilities of CTP systems for each reference bands in Figure \ref{fig:ctpdetectionrate}. Contours are the radius correction factor $X_R$ for CTPs with different spectral types.

    Most notably, the \emph{Kepler}-$K_s$ photometry can resolve the depth variation of CTPs with wide ranges of spectral type combinations. Although CTPs comprising stars with nearly identical spectral types ($T_\mathrm{eff,sec}/T_\mathrm{eff,pri}>0.9$) are hard to show detectable depth variations, the planet radius estimate from this CTP population is close to the true value so they do not plague the planet classifications as severely as the other CTPs do. 

    Using $z$ and TESS band as the reference band is not as effective in resolving the depth variations. The CTP system involving a K dwarf primary ($\sim 4500$K) and an M dwarf secondary ($\sim 3500$K) is most ameanble to display resolvable depth variations for \emph{Kepler}-$z$ and -TESS photoemtry, even in the case where $\sigma_\mathrm{z,TESS}/\sigma_\mathrm{KP}=10$. The adequate binary color-difference and the small size of the secondary star would allow larger amplitude of depth variations and render this CTP population the optimal targets for the search of depth-difference. 

    When $\sigma_\mathrm{z,TESS}/\sigma_\mathrm{KP}=3$, depth variations can be resolved for more than 40\% of the CTPs with temperature-ratio of 0.6-0.9. CTPs containing stars with very similar spectral types (i.e. two G dwarfs or two F dwarfs, etc.) or with very different spectral types (i.e. an F/G dwarf paired with an M dwarf) are still difficult to display detectable depth variations, which is due to the lack of color-difference and the heavy flux contamination of primaries (respectively), as explained above. When TESS and $z$ band can achieve a \emph{Kepler}-like photometric precision, one can expect to observe the depth variations of most CTPs, but only comparable to the most noisy $K_s$ photometry ($\sigma_\mathrm{K}/\sigma_\mathrm{KP}=10$).

    \subsubsection{Depth variation detection probability of CTPs with different planet radii}
    \label{sec:spec_exclusion_rate}

\begin{figure}[ht]
        \centering
        \includegraphics[width=0.6\textwidth]{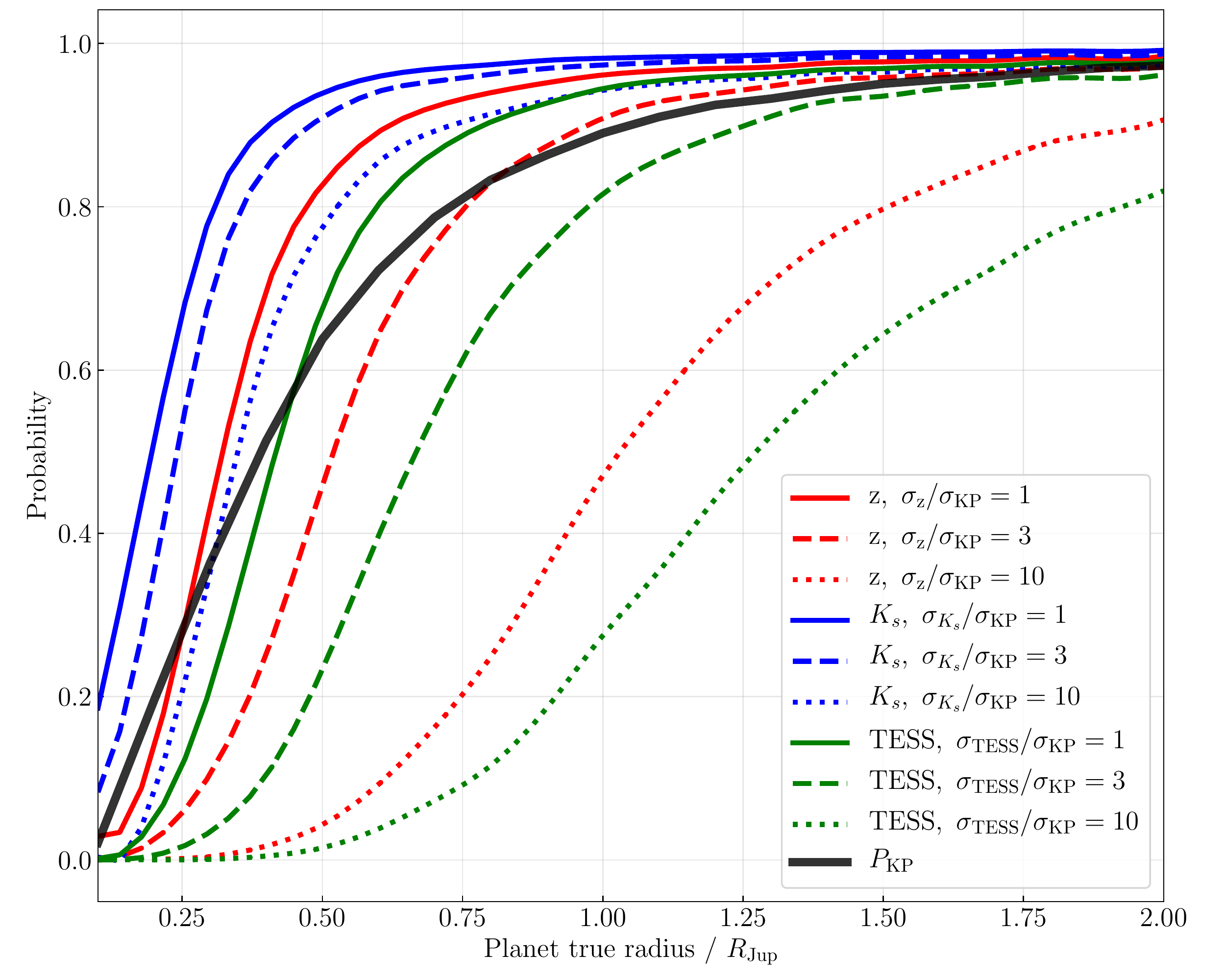}
        \caption{This figure presents the probability of a CTP system will show detectable double-band depth variation (colored lines) and the probability of a CTP system can show detectable transit signals on \emph{Kepler} band (black thick line), as a function of the true radius of resident planet.}
        \label{fig:ctp_true_prad_ruleout}
    \end{figure}

    We present a framework to calculate the overall exclusion rate of \emph{Kepler}-detected CTPs. We list the following key ingredients needed to compute the overall exclusion rate of \emph{Kepler}-detected false positives in double-band photometry. We start with the probability that the planet in CTP configuration having a radius of $R_p$, $P^\mathrm{CTP} (R_p)$. Only a fraction of CTPs with given planet radii can be still detectable on \emph{Kepler} band after the dilution. To obtain the modified planet radius distribution of \textbf{\emph{Kepler}-detected} CTPs, we derive the \emph{Kepler} detection probability for CTPs with given $R_p$, $P_\mathrm{KP}(R_p)$. This quantity is plotted as the black thick line in Figure \ref{fig:ctp_true_prad_ruleout}. More than 70 \% of giant planets ($R_p>0.6~R_\mathrm{jup}$) can still be detected by \emph{Kepler} band. However, the detection probability will decrease sharply for small planets as their apparent transit depths are susceptible to get diluted to be undetectable by the primary star. For those Earth-sized planets that occur in CTP systems, there is nearly zero chance for their diluted transits to be detected by \emph{Kepler} band.
    
    Next we calculate the specific exclusion rate for a \emph{Kepler}-detected CTP with given $R_p$, $f^\mathrm{CTP}_\mathrm{spec,ex} (R_p)$. This quantity can be obtained by integrating the detection probability shown in Figure \ref{fig:ctpdetectionrate} over the whole $T_\mathrm{pri}-T_\mathrm{sec}/T_\mathrm{pri}$ space. The expected specific exclusion rates computed with different reference bands and photometric precisions are displayed as the colored lines in Figure \ref{fig:ctp_true_prad_ruleout}. The specific exclusion rate increases with the radius of the resident planet since the amplitude of depth difference scales with planet size. Earth-size planets' specific exclusion rate is nearly zero for all reference bands. $K_s$ band has a specific exclusion rate larger than 80\% for all planets larger than $0.5 R_\mathrm{jup}$. The reference band photometric precision has a small impact on the $K_s$ band's specific exclusion rate while affecting the rate of $z$ and TESS more seriously. The specific exclusion rate for noisy ($\sigma_K/\sigma_\mathrm{KP}=10$) $K_s$ photometry is comparable to that of $z$ and TESS with best photometry precision ($\sigma_\mathrm{z,TESS}/\sigma_\mathrm{KP}=1$).

    Based on all the quantities, the overall exclusion rate, $f^\mathrm{CTP}_\mathrm{ex}$, can be computed. With selected planet radius function in CTPs, the probability density function of a \emph{Kepler}-detected CTP having a planet with radius $R_p$, $P^\mathrm{CTP}_\mathrm{KP}(R_p)$, can be written as
    
    \begin{equation}
    \label{eq:kpctp}
        P^\mathrm{CTP}_\mathrm{KP}(R_p) = N_0 P_\mathrm{KP}(R_p)  \times P^\mathrm{CTP}(R_p)
    \end{equation}
    
    $N_0$ is the nomalization factor. The overall exclusion rate can be computed by integrating the specific exclusion rate with Equation \ref{eq:kpctp} over $R_p$

    \begin{equation}
    f^\mathrm{CTP}_\mathrm{ex} = \int f^\mathrm{CTP}_\mathrm{spec,ex}(R_p) \times P^\mathrm{CTP}_\mathrm{KP}(R_p) d R_p 
    \end{equation}

\section{Constraining the false positive probability with double-band photometry}
\label{sec:reducefpp}

In the previous section, we take a clear cut in depth variation SNR and use this as the sole metric to identify BEBs and CTPs, and estimate their rule-out fraction. While a depth-variation detection limit of five is acceptable, it is desirable to evaluate the signal's status of being a false positive quantitatively. As the apparent depth double-band difference behaves differently between false positive and genuine planet transit, we could incorporate the double-band photometry into the false positive probability (FPP) computation, yielding a more improved and reliable result. 

This section aims to demonstrate the capability of multi-color photometry in constraining the FPP of background eclipsing binaries and companion transiting planets. The calculation of FPP makes use of the details of the candidate's light curve and quantify how well they agree with the expectance of each false positive scenario. We parameterize the double-band transit signals into only two observables: the apparent ``transit'' depths on \emph{Kepler} band and reference band. Rather than including all conceivable false positive scenarios in the FPP estimate, which is beyond the scope of this work, we present examples where only a certain FP scenario is most probable. As a result, the FPP calculation will be simplified while the results remain reasonable in showing the FPP's improvement realized by multi-color photometry. Adopting a Bayesian probabilistic approach, the FPP of a transit signal can be expressed as:

\begin{equation}
\label{eq:fpp}
       FPP = \frac{p_{FP} \mathcal{L}_\mathrm{FP}}{p_\mathrm{TP}\mathcal{L}_\mathrm{TP}+p_\mathrm{FP} \mathcal{L}_\mathrm{FP}}  
\end{equation}

where $p$ is the \emph{a priori} of the hypothesis, and $\mathcal{L}$ is the likelihood of the hypothesis evaluated by the observed ``transit'' depths in two bands, describing how well the depths obtained from double band photometry match those predicted from the corresponding hypothetical model. The subscript TP denotes the \textbf{true positive} hypothesis, which is usually the case of a genuine planet transit with the right radius estimate, and FP denotes the \textbf{false positive}. We will point out the definition of each hypothesis in the following case study.

\subsection{Background eclipsing binaries mimicking Earth-sized planets}
\label{sec:beb_fpp}

False positives involving an eclipsing binary can be categorized into three groups: undiluted eclipsing binaries (EBs), hierarchical triples (HTs), and blended background eclipsing binaries (BEBs), and each of them tend to produce different orders of eclipse depths, with EBs mimicking Jupiter-size transits, HTs mimicking Neptune-size transits and BEBs mimicking the sub-Neptune size and smaller planet transits \citep{Morton2012,Morton2016,Fressin2013}. We note that this statement is not definitive but empirical in the sense that heavy dilution will attenuate the eclipse depth shallow.

Driven by this inference, we evaluate the impact of multi-color photometry on the FPP by the example: an earth-sized planet transiting signal (100 ppm) found from a bright ($m_\mathrm{KP}=11$) G dwarf (one solar mass) located in low galactic latitude region in \emph{Kepler} FoV ($b\sim 7$ degree). The reason that we assign shallow depth and low-latitude region to the signal is that, under such circumstance and among all the possible false positive scenarios, background eclipsing binary is most likely to represent the \textbf{false positive} hypothesis in the FPP calculation \citep{Fressin2013,Batalha2010}. The \textbf{true positive} scenario, in this case, is a genuine Earth-size planet transiting the target star.

\subsubsection{FPP calculation}
 \label{sec:beb_lhood}
 
 According to equation \ref{eq:fpp}, we determined the prior and likelihood of BEB and planet hypothesis in the following manners:

 \emph{Prior of BEB}——The prior probability of BEB $\pi_\mathrm{BEB}$ should be the product of (1) the probability of a background star being blended within the aperture of the target and (2) the probability of the blended star being an eclipsing binary. The first term is set to be 0.37, derived from Figure 1 of \cite{Morton2011}; the second term is set to be 0.79\%, which is the occurrence rate of short-period eclipsing binaries discovered in \emph{Kepler} field \citep{Slawson2011}. We have $p_\mathrm{BEB}=0.003$.
 
 \emph{Prior of planet}——To obtain the prior probability of the signal being a genuine transiting Earth-size planet, we rely on the frequency of detected Earth-size planets in entire \emph{Kepler} survey. We count the number of confirmed Earth-size planets and candidates from KOI lists, then divide the number by the number of KIC stars, which yields $p_\mathrm{pl}=\frac{607}{156453}=0.004$.

 \emph{Likelihood of BEB}——To create the likelihood space of BEB, we generate background EBs from the TRILEGAL simulation following the same formula described in \ref{sec:bebsim}, except that all foreground stars are subject to be solar-like G dwarf stars (a gaussian mass distribution of $1 \pm 0.05 M_\odot$). Then we compute the diluted primary eclipse depths in \emph{Kepler} band and reference bands. The sample's logarithmic \emph{Kepler} depth $\log \delta_\mathrm{KP}$, and double-band depth ratio $\delta_\mathrm{RF}/\delta_\mathrm{KP}$ will be used to define the two-dimensional probability density function (PDF) of BEB hypothesis. The likelihood is then
    
 \begin{equation}
 \label{eq:beb_likelihood}
    \mathcal{L}_\mathrm{BEB} = \int PDF_\mathrm{sim}(\bm{\delta}) PDF_\mathrm{obs}(\bm{\delta}) d \bm{\delta}
 \end{equation}
 where $\bm{\delta}$ is the vector of $(\log \delta_\mathrm{KP},~\delta_\mathrm{RF}/\delta_\mathrm{KP})$, and $PDF_\mathrm{sim}$ is the posterior PDF mapped by BEB simulation, $PDF_\mathrm{obs}$ is the posterior PDF of observed signal, which is a gaussian distribution centered on $\bm{\delta}$, with depth uncertainty as its standard deviation. We define the observed $\delta_\mathrm{KP}$ at $100 \pm 20$ ppm, and leave the $\delta_\mathrm{RF}$ as a varying value. We set the depth uncertainties on the reference band, $\sigma_\mathrm{RF}$, to be 20 or 50 ppm. 

 \emph{Likelihood of planet}——For the planet hypothesis, we compute the planet radius based on the observed $\delta_\mathrm{KP}$ (within errors) and the simulated foreground star radius. Then we compute the apparent transit depth on reference bands with limb-darkened transit model \citep{2002ApJ...580L.171M}, with appropriate quadratic limb darkening coefficients taken from \cite{Claret2011} and \cite{Claret2017}. The likelihoods of planets is evaluated in the same way as for the BEBs.
    
\begin{figure}[ht]
    \centering
    \includegraphics[width=0.75\linewidth]{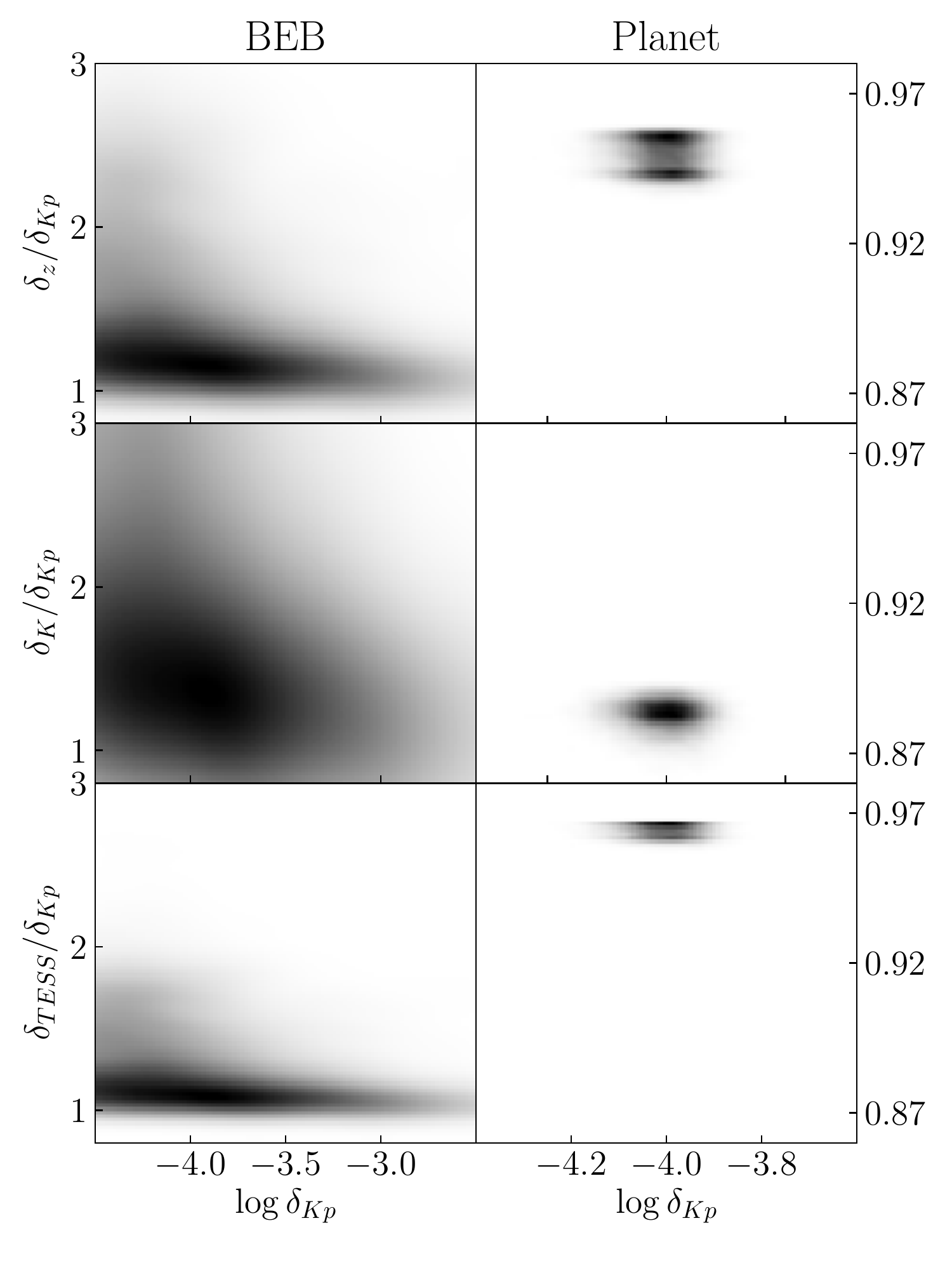}
    \caption{Likelihood space for BEB and planet scenarios described in Section \ref{sec:beb_fpp}. The area with deeper color means a higher probability that our hypothesis simulations have a certain value of $(\log \delta_\mathrm{KP},~\delta_\mathrm{RF}/\delta_\mathrm{KP})$ in the region. Note that the depth ratio for two hypotheses is different. BEB tends to show a larger apparent depth in the redder reference band compared to \emph{Kepler}. Planet transits show a smaller depth due to stellar limb darkening. The observed depth ratio for different photometry pair is separated at different levels, with \emph{Kepler}-$K_s$ pair showing the most distinctive depth ratio distribution for BEB and planet scenarios; while the \emph{Kepler}-TESS shows a similar depth ratio between the two hypotheses, in both cases the depth ratios are close to 1.}
    \label{fig:beb_fpp_example}
\end{figure}

 Figure \ref{fig:beb_fpp_example} shows an example of the PDFs of BEB and planet hypothesis. The expected depth ratio distribution of each hypothesis is quite different. BEBs tend to produce deeper $\delta_\mathrm{RF}$ than $\delta_\mathrm{KP}$ because the background binaries are generally of low masses and appear consequently red in color, while the foreground G dwarf is less luminous in longer wavelengths. Thus the flux contamination of foreground stars will be less significant on the redder reference bands. The depth ratio expected for planet transits is smaller than 1, which is due to the stellar limb darkening effect, which states that light with longer wavelengths is more evenly distributed on stellar disk than light with shorter wavelengths. Therefore the planet will block more blue light than red light, and the $\delta_\mathrm{RF}$ will be smaller than $\delta_\mathrm{KP}$. 
    
 Moreover, the most probable depth ratio expected for each hypothesis, represented by the area with the deepest color in Figure \ref{fig:beb_fpp_example}, is also unique when adopting different reference bands. \emph{Kepler}-$K_s$ pair has the most separated $\delta_\mathrm{RF}/\delta_\mathrm{KP}$ distributions for two hypotheses, with the most probable ratio value of 1.5 for BEB hypothesis and 0.88 for planet transits. \emph{Kepler}-TESS photometry has the least separable depth ratios for two hypotheses. The distinction between BEB's and the planet's likelihood distributions results from the photometry pair's capability to reveal the color change during the eclipse/transit, which is consistent with the pattern we demonstrate in Figure \ref{fig:beb_ruleout} and \ref{fig:BEB_plradius_ruleout}.
 
 \subsubsection{Improved FPP from double photometry}
 
\begin{figure}[ht] 
	\centering 
	
	\includegraphics[width=0.7\linewidth]{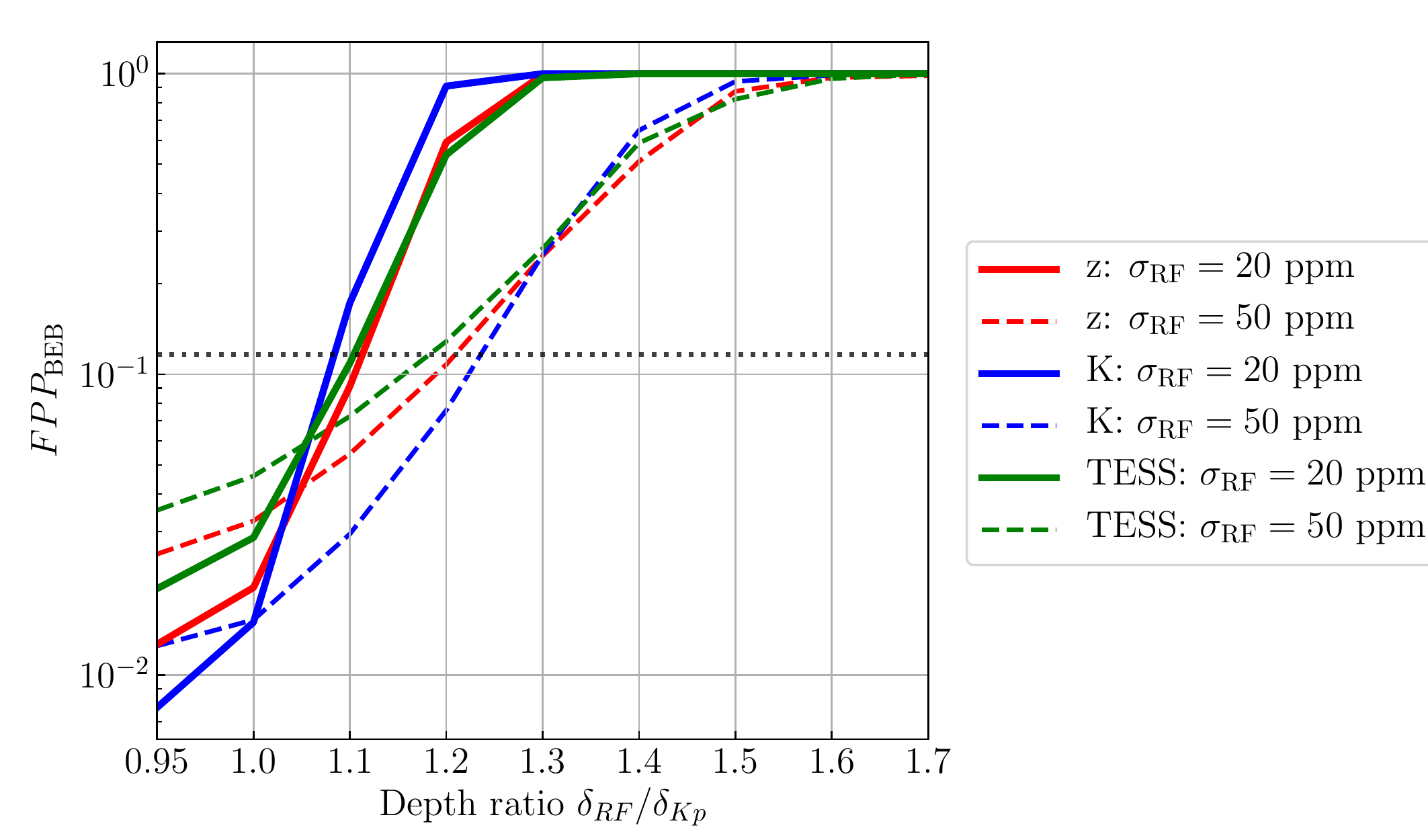}
	\caption{The false positive probability that an Earth-sized candidate is caused by a background eclipsing binary, evaluated by the observed double-band depth-ratio. The black dotted line is the single-band FPP derived from \emph{Kepler} photometry alone. The solid and dashed lines are cases when the depth observed by reference band is of different uncertainties. See Section \ref{sec:beb_fpp} for more discussion.}
	\label{fig:beb_fpp}
\end{figure}

    We present the improved FPP results in Figure \ref{fig:beb_fpp}, and compare to the single-band FPP. Besides selecting three different photometric bands as the reference bands, we also provide two cases when the depth uncertainty is different (represented by solid and dashed line). The general trend is that as the observed depth ratio increases, the FPP goes higher. When the observed depth ratio is close to unity, the FPP will decrease, supporting the planet hypothesis.
    
    It is also shown from the figure that the order of FPP increment or decrement varies across reference bands. $K_s$ band shows the best performance in discriminating the planets and FP scenarios. If $\delta_{K_s}/\delta_\mathrm{KP}=1$, the FPP will decrease by an order. Also, it takes a smaller depth ratio for the \emph{Kepler}-$K_s$ pair to fully confirm the false positive origin of the signal than the other two photometry pairs (1.3 and 1.7 for the two cases). This is because \emph{Kepler}-$K_s$ pair has the most discrepant $\delta_{K_s}/\delta_\mathrm{KP}$ distributions for planet and BEB hypothesis (see Figure \ref{fig:beb_fpp_example}), which makes it easier to distinguish the underlying hypothesized scenario that creates the observed depth ratio. 
    
    Secondly, the comparison between two cases with different depth uncertainties suggests that the actual value of depth uncertainty will affect the double-band FPP more significantly than the different choice of reference band. For example, if a depth ratio of 1.3 is observed, the corresponding FPPs derived from $\sigma_\mathrm{RF} = 50$ ppm are generally ten times lower than the case of 20 ppm, while FPPs are close to each other for all three photometry pairs at a given $\sigma_\mathrm{RF}$. A result of this effect is that larger uncertainties in depth measurement would require a larger observed depth ratio to falsify a signal. The critical depth ratio, which makes the FPPs increase to nearly 100\%, is 1.3-1.4 and 1.7-1.8 for $\sigma_\mathrm{RF} = 20$ ppm and $\sigma_\mathrm{RF} = 50$ ppm, respectively.
    
    Finally, we may compare the FPP results with the depth-difference identification method adopted in Section \ref{sec:ruleoutfp}. If we calculate the SNR$_\mathrm{RF-KP}$ for these critical depth-ratios, the value is only around 1.4, which is well below the threshold of 5 we set in Section \ref{sec:ruleoutfp}. This comparison suggests that incorporating the double-band photometry in FPP calculation can more easily identify false positives.

\subsection{Which is the planet-host star?}
\label{sec:ctpfpp}

Identifying the planet-host star in a binary system is crucial in correcting the dilution of observed transit signals. Applying the double-band photometry FPP calculation to the model of companion transiting planets provides a natural way to distinguish the planet-host star in a binary system. We illustrate this by providing a specific observation event. Suppose a Super-earth-size transit candidate ($\delta_\mathrm{KP} \sim 200$ ppm) is observed from a solar-like G dwarf, but it is unknown whether the target star is single or not. To predict the multi-band transit depths for each hypothesis, we simulate binary populations with a primary mass of around one solar mass and a varying secondary mass, according to the procedure described in Section \ref{sec:bebsim}. We assume that, before the transit on reference band is obtained, the apparent depth observed by \emph{Kepler} alone provides no clues about the host star of the planet; that is, the planet is likely to orbit either star in the binary with equal probability, hence $\pi_\mathrm{TP}=\pi_\mathrm{FP}=50\%$. Here we define the \textbf{true positive} as the planet orbiting the brighter target star, while \textbf{false positive} as the planet orbiting the fainter companion of the target star. 

To determine the likelihood of each hypothesis, we first draw the apparent $\delta_\mathrm{KP}$ from a Gaussian distribution with a mean value of 200 ppm and a standard deviation of 20 ppm; the latter quantity is the depth uncertainty $\sigma_\mathrm{KP}$ we assigned to the \emph{Kepler} photometry. Then we recover the true planet radius from the apparent depth $\delta_\mathrm{KP}$ for each instance, assuming the primary/secondary star is the planet-host star. We reject those instances with a recovered planet radius larger than 22 $R_\oplus$ to avoid overestimation from systems that unlikely exists. Then we calculate the observed depth on reference band $\delta_\mathrm{RF}$ and establish the PDFs. Finally, we evaluate the likelihoods in the similar manner described in Section \ref{sec:beb_lhood}.

\begin{figure}[ht]
    \centering
    \includegraphics[width=0.7\textwidth]{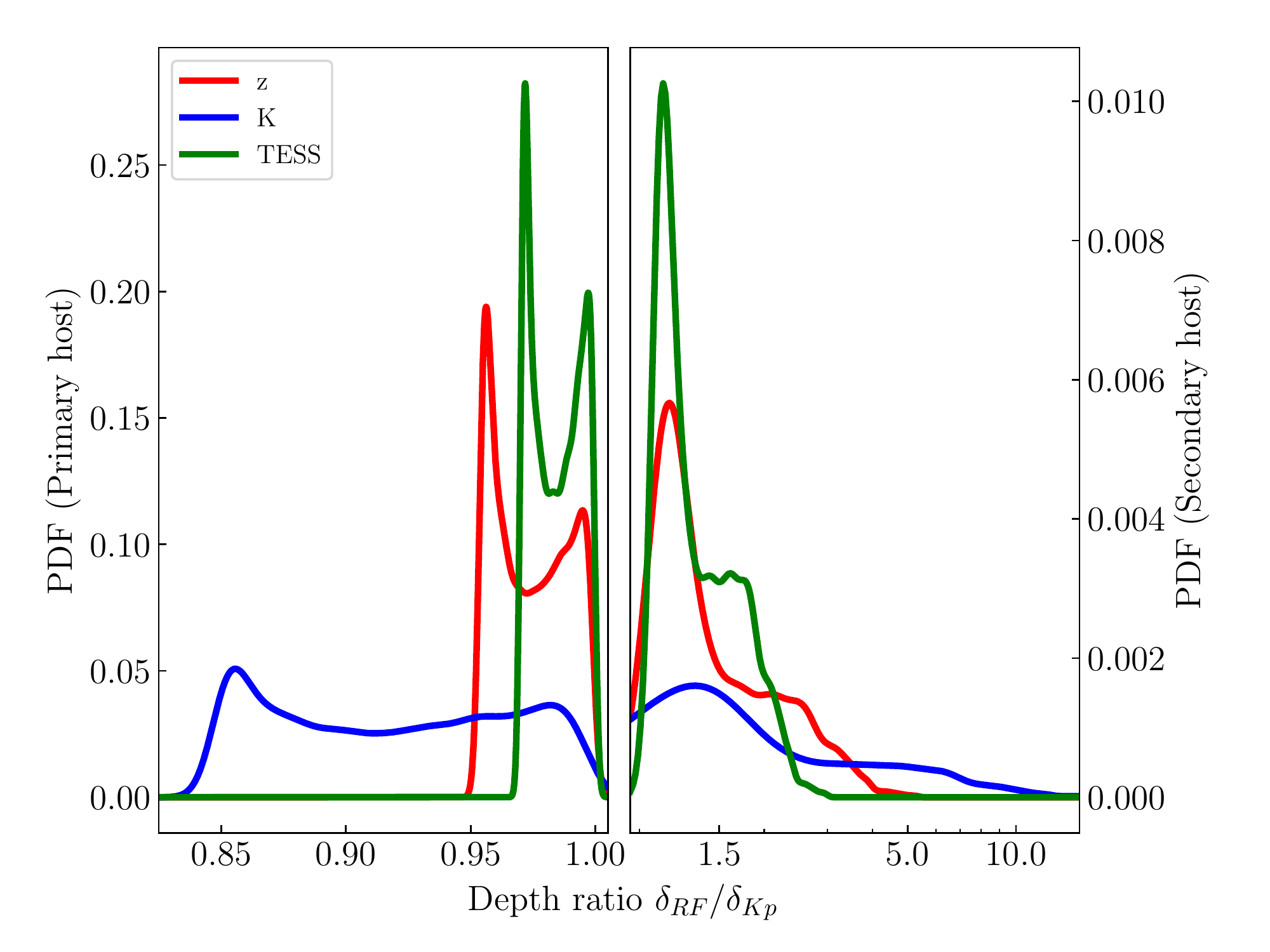}
    \caption{The likelihood distribution of the transit depth if an apparent super-earth-size ($\delta_\mathrm{KP} \sim 200$ppm) candidate is observed on three reference bands, under the assumptions that the planet is transiting the primary star (left) or the secondary star (right).  }
    \label{fig:ctp_lhood}
\end{figure}

We present the likelihood distribution of the two hypotheses in Figure \ref{fig:ctp_lhood}. If the primary star is the host star, the apparent radius of the planet is close to the true value, and the depth observed on different photometry band will not differ a lot. This is because the primary star usually makes up most total photometry. Therefore the variance of fainter companion's contamination on different photometry bands would be relatively small. The situation can be reversed when the secondary star is the host star, which is shown in the right panel of Figure \ref{fig:ctp_lhood} that the apparent transit depth observed in the reference bands can be up to 5-10 times of that observed on \emph{Kepler} band. This is because the true planet radius is usually much larger than the apparent radius so that the depth variation will increase, too. The apparent depth on the $K_s$ band can vary in a wide range, from 170 ppm to 2000 ppm. On the other hand, TESS and $z$ band expect a much narrower possible $\delta_\mathrm{RF}/\delta_\mathrm{KP}$ range. 

\begin{figure}[ht]
    \centering
    \includegraphics[width=0.7\textwidth]{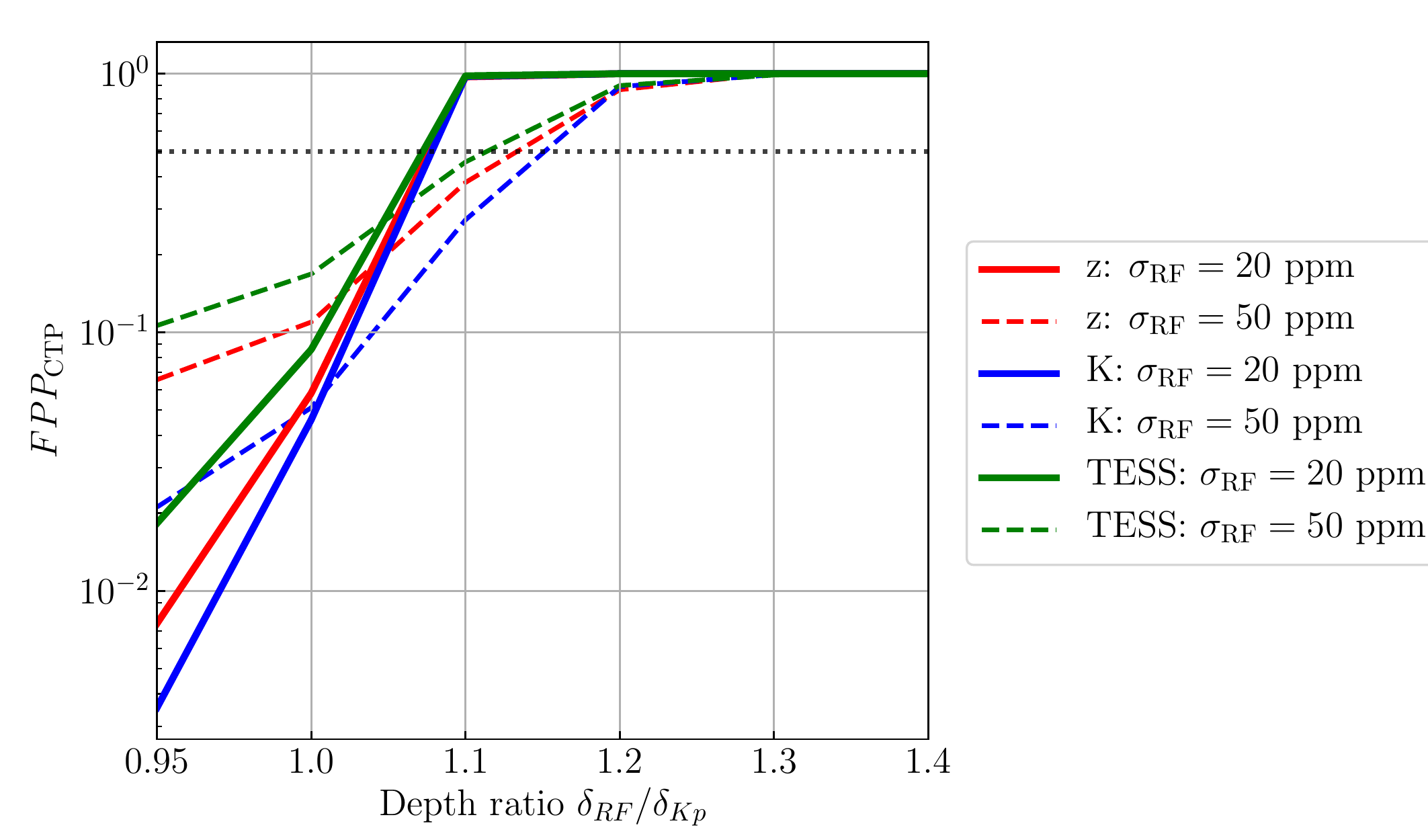}
    \caption{The false positive probability that a super-earth-size planet candidate is transiting the unseen stellar companion of the solar-like target star, evaluated by double-band depth-ratio. The black dotted line represents the uninformative prior probability that the planet is equally likely to orbit either of the stars in the binary system inferred from one-band photometry (prior = 50\% for each star). The solid and dashed lines are cases when the depth observed by the reference band is of different uncertainties. See Section \ref{sec:ctpfpp} for more discussion.}
    \label{fig:ctp_fpp}
\end{figure}

We calculate the probability of planet host star based on a set of transit depth ratio values and present the result in Figure \ref{fig:ctp_fpp}. We also provide two cases with different $\sigma_\mathrm{RF}$. As expected, the depth ratios close to unity will decrease the FPP, which suggests that the target star is indeed the planet host star. FPPs increase with the non-unity value of depth ratios, which alerts that the planet may be transiting the secondary companion of the target star. Larger depth uncertainties from the reference bands will lead to a loosely constrained FPP value. We find the SNR$_\mathrm{RF-KP}$ of the critical depth ratios to be 0.7 and 1.2, for two cases, respectively.

\section{Discussion}
\label{sec:discussion}
\subsection{Comparing two multi-color approaches to identifying false positives}

In this work, we demonstrate two methods by which we can use the color-dependence feature of transit candidates to reveal and rule out false positives. The first method is illustrated in section \ref{sec:ruleoutfp}, where we measure the SNR of the difference between observed depths and identify those false positives whose SNR of depth-variation is large enough to confidently confirm that the observed transit depth in two photometry is statistically inconsistent, thereby pointing out the possible presence of blending stars or the chance-alignment of eclipsing binary. In section \ref{sec:reducefpp} we adopt the second method. We compare the observed depth ratio with a comprehensive set of depth-ratios simulated under different hypotheses on the observed transit signal's origin, and we adopt a Bayesian approach to calculate the false positive probability of the signal based on double-band photometry.

The underlying assumptions for these two methods are a bit different. The basic assumption for the first method is that the apparent depth of false positives should be different. To identify this difference, we set the detection threshold SNR$_\mathrm{RF-KP}$ as 5 to evaluate whether the difference is real or due to statistical fluctuation. With this said, satisfying this detection criterion (SNR$_\mathrm{RF-KP} > 5$ and SNR$_\mathrm{KP}>7$) will require the signal to be observed with sufficient high SNR on reference band (SNR$_\mathrm{RF} > 12$), which is not only determined by the color-difference between the blends and targets, but also by the reference band's relative effective wavelength compared to the main photometry, and the photometric precision thereof. The approach is widely adopted in the early ground-based wide-field survey to reject transit signals \citep{ODonovan2006, Colon2012}, but is mainly subjected to vetting the transit signals' with deep depths ($\sim 1\%$) for which sufficient SNR can be attained. Small planet transit candidates may not be the optimal targets for multi-color vetting; as suggested by Figure \ref{fig:BEB_plradius_ruleout}, the detectability of small planet candidate's depth variation is most susceptible to the choice of reference band and precision. Thus, although we could identify blends and exclude them based only on one such metric, the definitive threshold will implicitly demand high SNR on the signal detection by reference photometry. The expected yield of false positive identification is strongly limited by the specifics of reference band and noise systematics.

The basic assumption for the FPP calculation is that the double-band depth ratio of planet transit is different from that of false positives. In simpler terms, we could falsify a transit signal as long as we can tell it does not resemble the planet transit's light curves. This is why we find in Section \ref{sec:reducefpp} that the critical depth ratio's corresponding SNR$_\mathrm{RF-KP}$ is usually much smaller than 5 since the observed depth ratio cannot be reproduced by most of the hypothetical planet-transit systems. However, since we adopt a Bayesian approach to achieve this goal, the value of double-band FPP will indeed vary as we change each hypothesis's \emph{a priori} probability and as we change the way we simulate hypothesis likelihood PDFs. If the \emph{a priori} probability of the planet hypothesis is larger, the corresponding FPP will be generally lower than before.

Nevertheless, we note that this is the common feature of Bayesian probability, which strongly depends on the models and assumptions. One advantage of double-band FPP, which we did not explicitly address in this work, is that it can work in parallel with other observational constraints to explore and narrow down the possible blending systems that can produce the double-band light curves, e.g., high-resolution imaging can reveal both physical companions in wider orbits and line-of-sight objects. The dilution information derived from magnitude difference helps to locate the star that harbours the observed signal \citep{Furlan2017,Ziegler2016}. For targets including several stars, which are hardly resolved, joint analysis of multi-color transit light curves and SED of the targets may provide stronger characterization of the multiplicity and configuration of the system \citep{Miyakawa2021}.

\subsection{The role of limb darkening}

Limb darkening will also cause the planet transit's depth to vary on different bands. This is because fluxes with different wavelengths distribute on stellar disk differently, with the center of the disk appear bluer and the limb appears redder. As a consequence, the transit depths observed on bluer photometry will be deeper than redder photometry, which projects $\delta_\mathrm{RF}/\delta_\mathrm{KP}<1$ since our considered reference photometry are all redder than \emph{KP}. 

To this end, one may question that whether using SNR$_\mathrm{RF-KP}>5$ as the sole metric to reject blends would ambiguously exclude the cases of limb-darkened planet transits signals. However, for the false positives considered in this paper the apparent depth ratio would be $\delta_\mathrm{RF}/\delta_\mathrm{KP}>1$. According to our definition of BEB and CTP in \ref{sec:fpsim}, the \emph{blending components}, which generate the eclipses/transits, are often redder in color compared to their contaminant (stars from \emph{target population}) and thus the apparent depth would be deeper on red reference photometry. Therefore, we could discriminate blends from limb-darkened transits by comparing the apparent depth ratio.

One exception would be grazing giant planets for $b>0.8$ \citep{Tingley2004}, which only pass through the limb of star and will block more blue light than red light. In this case the apparent depth ratio would be $\delta_\mathrm{RF}/\delta_\mathrm{KP}>1$, resembling the chromaticity of stellar blends. However, high-impact-parameter transits are rather rare cases and we do not consider it when using color-dependent feature to reject blends.

\subsection{Impact of different stellar properties from KIC}
\label{sec:diff_prop_kic}

Though our \emph{target star} population mainly consists of G-type stars as KIC included, we were unable to fully recreate the KIC's stellar mass and  effective temperature distributions in our simulation (as shown in Figure \ref{Fig1:m_kp_teff_compare}). These three parameters, as demonstrated in Section \ref{sec:ruleoutfp}, are the major stellar properties that determine the detectability of double-band depth variations. The deviation between the simulation and KIC stars partly stems from the fact that we employed a more physical model (see Section \ref{sec:fpsim_detail}) to construct the stellar population, while KIC is a specially curated catalog following some selection rules.

As for this deviation's impact on the final results. Qualitatively speaking,  our target star population is on average hotter than KIC stars. In the BEB model, hotter foreground stars tend to produce earth-like transit signals, because they are brighter so the dilution is heavy, therefore the rule-out rate of BEB signals (summarized in Figure \ref{fig:BEB_plradius_ruleout}) is affected the most for earth-size BEBs, the other sizes of BEB signals only being minorly affected. The excess of BEBs with hotter foreground stars leads to an overestimation of earth-like BEB rule-out rate. As for the overall rule-out rate (summarized in Figure 4\&7), the case for $\sigma_\mathrm{RF} = \sigma_\mathrm{KP}$ is affected the most, while for other reference band noise levels, the identifiable earth-like BEBs only make up a small fraction of the total rule-out rate and thus are not affected much.

For CTP model, we demonstrate that CTPs with hot primaries are the hardest to rule out at low reference band precision, espeacially for $z$ and TESS band (see Figure \ref{fig:ctpdetectionrate}). Thus, we expect the excess of hotter stars in our sample will lead to an underestimation on the rule-out probabilities given planet radius (summarized in Figure \ref{fig:ctp_true_prad_ruleout}) for $z$ and TESS band in $\sigma_\mathrm{RF}/\sigma_\mathrm{KP}=3,~10$.

However, we do manage to construct a KIC-like target star simulation to quantitatively measure the uncertainties in final results, even though we adopt some unphysical input distributions. The rule-out probabilities in both BEB and CTP models vary within 10\%, even for the most severely affected cases.

\section{Summary}
\label{sec:summary}

In this work, we study how many false positives can be identified in terms of their chromatic depth feature if another reference band different from \emph{Kepler} band were added and observing with \emph{Kepler}. We mainly focus on two types of false positives. One is that when the target star is blended with a pair background eclipsing binary (BEB); another is when the planet is transiting the unresolved stellar companion to the target star (we term this as companion transiting planet, or as CTP). We build up physics-based models for these two false positives. To systematically estimate what fraction of BEBs and CTPs can be identified through their color-dependent feature, we simulate representative BEB and CTP populations according to \emph{Kepler} observations and galactic star population models. We derive their multi-band photometry based on isochrone interpolation and bolometric correction. We calculate the signal-to-noise ratio of the difference of observed transit depth, SNR$_\mathrm{RF-KP}$, observed between \emph{Kepler} and reference bands, which include  $z$, $K_s$, and TESS. We identify the subset of false positives whose SNR$_\mathrm{RF-KP}$ is larger than five as the fraction of false positives that can be readily ruled out and show how this fraction varies with different levels of photometric precision we assigned to the reference bands. Some false positives cannot gain sufficient high SNR$_\mathrm{RF-KP}$. However, since the depth variation predicted by false positive hypothesis and planet hypothesis is quantitatively different, our multi-color model enabled us to calculate the transit signals' false positive probability based on double-band photometry. We summarize the conclusion of this work as follows.

\begin{itemize}

    \item Background eclipsing binaries (BEBs) can be effectively ruled out by combined identification of double-band depth difference (Equation \ref{eq:depthdifference}) and primary and secondary difference (Equation \ref{eq:beb_prisec_diff}); We list the overall rule-out fraction via identifying the depth-difference of BEB in Figure \ref{fig:BEB_plradius_ruleout}. $K_s$, $z$, and TESS, when adopted as the reference band to \emph{Kepler}, can rule out 95\%, 80\%, and 65\% of \emph{Kepler} candidates mimicked by BEBs if both of the photometry can attain a precision like \emph{Kepler} 's. Assuming a larger photometry precision of the reference band, the overall rule-out fraction of TESS and $z$ band will drop drastically towards around 10\%. The sharp decrease is mainly due to the undetectable depth variation of shallow-depth candidates that comprise most BEB signals, and only those signals with deep depths can show detectable depth variations. However, \emph{Kepler}$-K_s$ dual photometry is only slightly affected even if it is assigned with the largest noise, i.e. ten times of Kepler-like precision, reaching 70\% which is generally better than the TESS and $z$ band.·
    
    \item The amplitude of the depth variations of CTPs comes from the true planet radius and the color-difference between the binary stars the planet resides. We calculate the detection probability for \emph{Kepler}-detected CTPs with certain spectral type binary members (Figure \ref{fig:ctpdetectionrate}). The CTPs that consist of the smallest stars, such as a K dwarf and an M dwarf, often show detectable depth variations, while equal-temperature CTPs are the hardest to identify for all the considered reference bands. Next, we present the detection probability of CTPs with given true planet radius, averaged over all possible CTP binary masses (Figure \ref{fig:ctp_true_prad_ruleout}). Using the provided detection probability curve, one can obtain the overall exclusion rate of CTPs expected in a \emph{Kepler}-like survey by integrating the radius distribution of planets transiting the secondary star in a binary system.

    \item We adopt a Bayesian approach to calculate the false positive probability (FPP) based on the observed depth ratio between two photometry observations. We apply the framework to identify whether an Earth-size candidate is a diluted signal from a background eclipsing binary. For given transit SNR on the reference band, $K_s$ band can better constrain the FPP of signal better than $z$ and TESS band. However, the photometric precision of the reference band would influence the degree of FPP improvement more than the specific choice of reference bands. Nevertheless, using double-band photometry to determine the FPP and reveal blends usually takes less demand on photometry precision compared to identifying the blends via SNR$_\mathrm{RF-KP}>5$ criterion. (Figure \ref{fig:beb_fpp}). 
    
    \item Applying the FPP calculation to our CTP model can assist in figuring out whether an Earth-size transit signal is a larger planet transiting the bound companion of the target star, or equivalently identifying the planet host star in the binary system. For this purpose, assuming that the planet is equal-likely to orbit either of the stars inferred from initial single-band observation, the probability of the primary host can be up to 99\% if the observed depth ratio is nearly unity. The confirmation of the secondary host will need the observed SNR$_\mathrm{RF-KP}$ to be at least 0.8 (Figure \ref{fig:ctp_fpp}).
    
\end{itemize}

With a more flexible treatment of input assumptions about field stars' properties, the simulation procedure we have developed can have broader applications to other surveys as long as we have the priors and models specific to the surveyed area, whether physical or empirical. One possible application is to use in the follow-up of TESS, as the pixel scale of TESS (21") is much larger than that of \emph{Kepler} (4"); thus, a considerable amount of blended stars will be imaged. Merited for its economical cost in a single observation, multi-color photometry can extensively search for clues of false positive in \emph{Kepler} if future missions employ it, and the valuable time of advanced telescopes would be spared.

\acknowledgments

This work is supported by the National Natural Science Foundation of China (Grant No. 11973028, 11933001, 1803012), and the National Key R\&D Program of China (2019YFA0706601). We also acknowledge the science research grants from the China Manned Space Project with NO.CMS-CSST-2021-A1.

\software{Astropy \citep{2018AJ....156..123A}, Matplotlib \citep{Hunter:2007}, Pandas \citep{reback2020pandas}, isochrones, \citep{2015ascl.soft03010M}, scipy \citep{mckinney-proc-scipy-2010}}

%\bibliography{mybib}
%\bibliographystyle{aasjournal}



\end{document}